\documentclass[10pt,twoside]{article}
\usepackage {amsmath, amssymb, amsthm, mathrsfs, amsfonts, bbm, ctable,graphicx,color,epsfig,stmaryrd,textpos} 

\usepackage[abs]{overpic}

\usepackage{fancyhdr}
\renewcommand{\subsectionmark}[1]{}
\newlength\Li \newlength\Lii
\newcommand{\sech}{\mbox{{sech}}}

\newcommand{\vdiv}{\mbox{{div}}}

\newcommand{\Spec}{\mbox{{Spec}}}

\date{}

\makeatletter
\def\@seccntformat#1{}
\makeatother
\numberwithin{equation}{section}
\renewcommand{\numberline}[1]{}

\makeatletter
\def\blfootnote{\xdef\@thefnmark{}\@footnotetext}
\makeatother

\title{Quantum Hydrodynamics with Trajectories: The Nonlinear Conservation Form Mixed/Discontinuous Galerkin Method with Applications in Chemistry} 
\author{C.~Michoski\textsuperscript{\dag}, \ J.A.~Evans\textsuperscript{*}, \ P.G.~Schmitz\textsuperscript{\ddag} \ \& \ A.~Vasseur\textsuperscript{**} \\ \\ \small{Departments of Mathematics}, \\ \small{Computational and Applied Mathematics,} \\ \small{Chemistry and Biochemistry} \\ \small{University of Texas at Austin}}

\usepackage[square,comma,numbers,sort&compress]{natbib}
\usepackage[pdftex,bookmarks]{hyperref}
\usepackage{hypernat}

\begin{document}
\maketitle
\begin{abstract}  We present a solution to the conservation form (Eulerian form) of the quantum hydrodynamic equations which arise in chemical dynamics by implementing a mixed/discontinuous Galerkin (MDG) finite element numerical scheme.  We show that this methodology is stable, showing good accuracy and a remarkable scale invariance in its solution space.  In addition the MDG method is robust, adapting well to various initial-boundary value problems of particular significance in a range of physical and chemical applications.  We further show explicitly how to recover the Lagrangian frame (or pathline) solutions.
\end{abstract}

\blfootnote{\textsuperscript\textdagger {\it michoski@cm.utexas.edu}, Department of Chemistry and Biochemistry}\blfootnote{*{\it evans@ices.utexas.edu}. Computational and Applied Mathematics}\blfootnote{\textsuperscript\ddag {\it pschmitz@math.utexas.edu}, Department of Mathematics} \blfootnote{**{\it vasseur@math.utexas.edu}, Department of Mathematics}

\tableofcontents

\section{\texorpdfstring{\protect\centering $\S 1$ Introduction}{\S 1 Introduction}}

Quantum hydrodynamics (QHD) has engendered substantial activity in the field of theoretical chemical dynamics, where one may refer to Wyatt et al.(\cite{Wyatt}) for a comprehensive introductory overview of the numerous recent results emerging from this blossoming field. 

The basic idea emerging from quantum chemistry in the context of QHD is to employ the time-dependent Schr\"odinger equation (TDSE) to solve for the dynamical properties (probability densities, ``particle'' velocities, etc.) of chemical systems.  In the same spirit in which the de Broglie-Bohm interpretation (see \cite{Holland,Bohm,Bohm2}) of quantum mechanics may be used to recover ``trajectories'' of individual fluid elements along the characteristics of motion of the solution, the QHD equations of Madelung and Bohm are derived as formally equivalent to the TDSE and thus comprise an alternative route to solutions which generate quantum trajectories that follow particles along their respective paths (see \cite{Wyatt} and \cite{Jungel} for a comprehensive overview).

\begin{figure}[h]
\centering
\includegraphics[width=10cm]{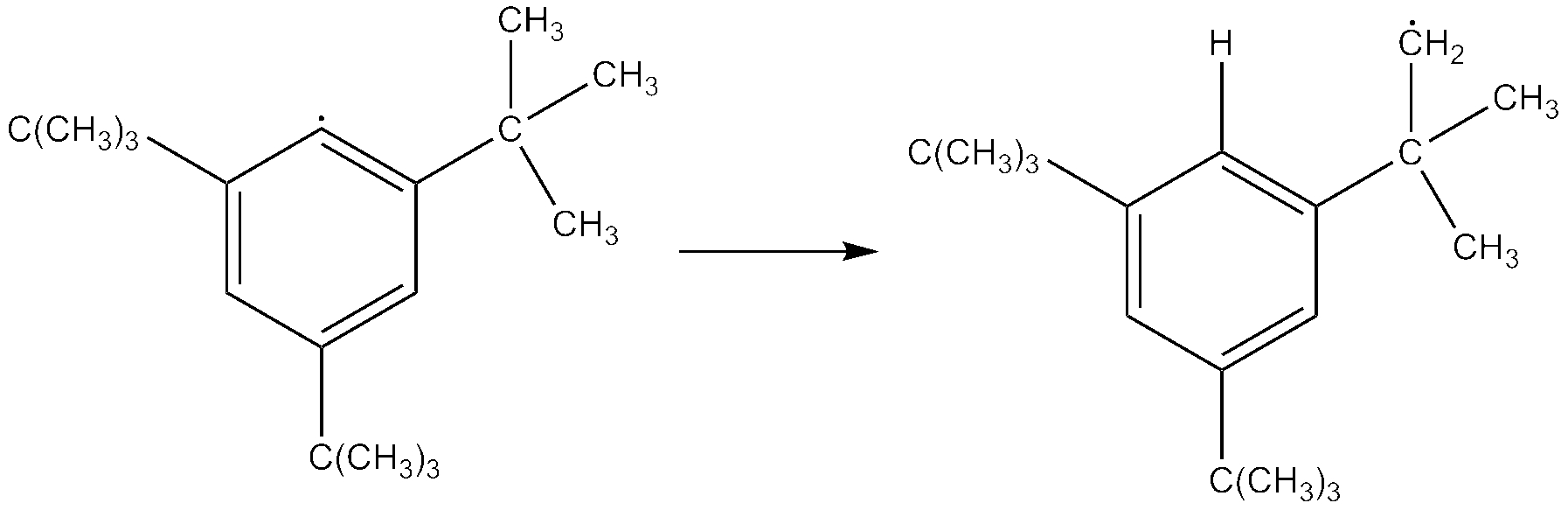}
\caption{Here we have the intramolecular rearrangement of the aryl radical $2,4,6$-tri-\emph{tert}-butylephenyl to $3,5$-di-\emph{tert}-butylneophyl (see \cite{BGBI} for details).}
\label{fig:proton}  
\end{figure}

\begin{figure}[!t]
\centering
\includegraphics[width=11cm]{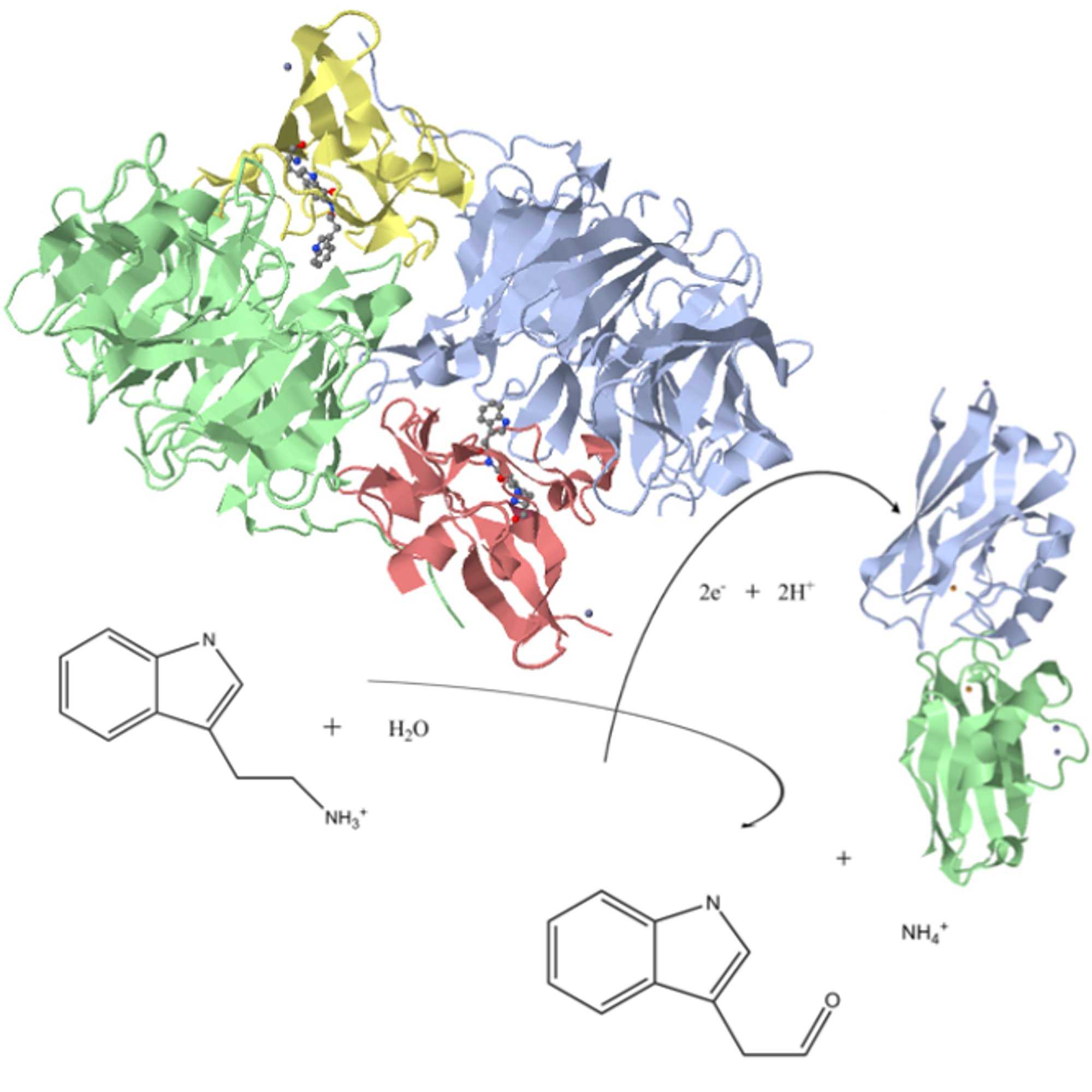}
\caption{Here we show an enzymatic catalysis -- an aromatic amine dehydrogenase (AADH) with a tryptophan tryptophyl quinone (TTQ) prosthetic group catalyzing the oxidative deamination of tryptamine  with an electron transfer to an arsenate reductase enzyme (see \cite{BGBI} and \cite{MRJHBRMSSL} for details, PDB codes: 1nwp (azurin), 2agy (AADH)).}
\label{fig:enzyme}  
\end{figure}

These solutions hold particular significance, where, in the context of the QHD formulation, it is possible to resolve the chemical dynamics of a vast number of reaction mechanisms known to have pathways dominated by quantum tunneling regimes.  Some of these systems include proton transfer reactions (for example see figure \ref{fig:proton}), conformational inversions, biologically important redox reactions in enzymatic catalysis reactions (see figure \ref{fig:enzyme}), and proton-coupled electron transfer reactions (refer to \cite{McMahon} and \cite{MRJHBRMSSL}). It is not yet clear if these types of methods may also have application at higher energies, for example in the halo nuclei tunneling occurring in fusion reactions (as seen, for example, in \cite{IYNU}).

Substantial research has been done in quantum hydrodynamics to find the best and fastest computational methodology for solving this system of equations.  In the standard methodology presented using the quantum trajectory method (QTM), for example, solutions to the QHD equations are found by transforming the system of equations, which is generally posited in the Eulerian fixed coordinate framework (see \cite{Madelung,Jungel,Gardner,JG}), into the same set of equations in the Lagrangian coordinate framework, which effectively follows solutions along particle trajectories; or along so-called ``Bohmian trajectories.''  The transformation from the Eulerian to the Lagrangian frame leads to a set of coupled equations which solve for two unknowns: the \emph{quantum action} $S(t,\vec{r})$ and the probability density or \emph{quantum amplitude} $\sqrt{\varrho(t,\vec{r})}=R(t,\vec{r})$ along the trajectories $\vec{r}(t,\boldsymbol{x})$ (e.g. see \cite{Wyatt} box 1.2).  The obvious advantage of the Lagrangian framework is reduced computational times, since solutions are only computed along a set of chosen trajectories; while clearly the disadvantage is the possibility of obscuring structure hidden within the continuum of the full solution, which may only emerge properly in convergent numerical schemes, and also the increased complications of transposing into more complicated settings: such as with functional or time dependencies on the potential term $V$, or including dissipative or rotational vector fields.

In addition, the numerical solutions to the above mentioned Lagrangian formulations have demonstrated characteristic behaviors which introduce certain technical difficulties at the level of formal analysis.  First, the system of equations are \emph{stiff}, which is to say, solutions to the system may locally or globally vary rapidly enough to become numerical unstable without reducing numerically to extremely small timesteps. Furthermore, there exists the so-called ``node problem,'' which is characterized by singularity formation (see \cite{Wyatt} for characterization of node types) along particle trajectories.  Another issue which arises is obtaining unique solutions, since there is not a unique choice of trajectories in the Lagrangian formulation (see for example \textsection{6} and appendix A).  And finally, boundary data is often treated without regard to the (often substantial) numerical residuals introduced in the weak entropy case, or taking into account consistency between the TDSE and the QHD system of equations (see for example \cite{MSEV} and \textsection{3}).

We introduce an alternative formulation to the standard solutions described above in $\varrho$ and $S$ and tracked with respect to the Lagrangian coordinate frame which is motivated by work of Gardner, Cockburn, et al. (see \cite{Gardner,Gardner2,Gardner3}).  Instead, we keep the system in its conservation form (instead of in a primitive variable form) in the Eulerian coordinate system (see \cite{Madelung}), and solve for the density $\varrho = \varrho(t,\boldsymbol{x})$ and the particle \emph{velocity} $\boldsymbol{v} = \boldsymbol{v}(t,\boldsymbol{x})$ (instead of the \emph{quantum action} $S$).  We show that these solutions may be used to easily recover the variables $S$ and $\psi$ in a single step; and may with little difficulty be transformed into their Lagrangian coordinate frame counterpart solutions $\varrho(t,\vec{r}),\boldsymbol{v}(t,\vec{r}),S(t,\vec{r})$ and $\psi(t,\vec{r})$, using the conservation equation (continuity equation), or by solving for pathlines in the sense of classical mechanics, or by any number of alternative so-called ``offset methods.''  Additionally, our solutions demonstrate a type of resolution invariance, which is to say that the behavior of our solutions are qualitatively equivalent at varying spatial resolutions, and compare favorably with solutions to the formally equivalent TDSE.  As a consequence, our conservation-based formulation is computationally competitive with Lagrangian formulations, up to a type of ``formal accuracy'' in the trajectory solutions.

Our solutions, as the Lagrangian formulated solutions mentioned above, still demonstrate a \emph{stiff} behavior.  However, also as the Lagrangian solutions above, and similarly to the classical CFL condition in fluid mechanics, we consider this a prohibitive but not insurmountable computational difficulty.  On the other hand, our solutions to the conservation form of QHD do not demonstrate the node problem (at least on Gaussian wavepackets) as expected, as the only type of node our formulation exhibits is for $\varrho\equiv 0$, which never occurs if we add a numerical ambient density $\varrho_{A}$ to the initial density $\rho_{|t=0}$.  The solution is stable when the ambient density is set to $\sim 11$ orders of magnitude smaller than $\max_{\Omega}(\varrho)$ over a computational domain $\Omega$.  We maintain that the addition of $\varrho_{A}$ to the initial density does not significantly change the numerical solution of the system of partial differential equations, while introducing the substantial benefit of significantly improving its stability.  Again, this behavior compares favorably with solutions to the TDSE, which also do not demonstrate the node problem.  On the other hand, computing solutions in the Lagrangian frame still offers substantial computational efficiency when compared to those in the Eulerian frame; due simply to relative density of solutions.

We begin in \textsection{2} by presenting the governing equations, then rescaling these equations in time for substantial improvement of numerical tractability.   Next we present the details of a computationally well-posed finite element discretization scheme leading to our approximate (numerical) solution. The scheme is based on a discontinuous Galerkin method for the QHD conservation laws and a mixed finite element method for the Bohmian quantum potential, which is inspired by \cite{Gardner4}.  In \textsection{3} we briefly derive the basic equations, and discuss the rather strong dependence on the formal and numerical equivalencies in the boundary data.  In \textsection{4} we derive an analytic test case which allows us to find the relative error in the discontinuous Galerkin mixed method, which shows that our formulation is near to numerically exact everywhere but at the boundaries (which is expected).  We proceed in \textsection{5} by testing the standard case of a hydrogen atom tunneling through an Eckart potential barrier, compare these results to a finite difference scheme for the TDSE, and then show how to use the continuity equation to recover the Lagrangian, or Bohmian, trajectories.  Next, in \textsection{6}, we show how to compute pathlines, recover the variables $\rho,\boldsymbol{u},\psi$ and  $S$ in both the Eulerian and Lagrangian frames, and compare the way in which these solutions relate to each other. We finish with some concluding remarks in \textsection{7}.

\section{\texorpdfstring{\protect\centering $\S 2$ Conservation Formulation of Quantum Hydrodynamics}{\S 2 Conservation Formulation of Quantum Hydrodynamics}}

Consider the following system of equations for $(s,x)\in T_{s}\times\Omega$, motivated by \cite{Wyatt}, where we have transformed  the solution space from the usual Lagrangian coordinate frame into the conservation form of the Eulerian coordinate frame:

\begin{align} \label{mass}& \partial_{s} \varrho + \nabla_{x}\cdot (\varrho \boldsymbol{v}) = 0, \\ \label{momentum}& \partial_{s}(\varrho m \boldsymbol{v}) + \nabla_{x}\cdot\Pi + \varrho\nabla_{x}V= 0,  \end{align} with initial conditions \[\varrho_{s=0}= \varrho_{0},\quad\mathrm{and}\quad\boldsymbol{v}_{s=0}=\boldsymbol{v}_{0}\] where $\varrho=\varrho(s,\boldsymbol{x})$ is the probability density corresponding to conservation equation (\ref{mass}), and $\boldsymbol{v}=\boldsymbol{v}(s,\boldsymbol{x})$ is the volume velocity corresponding to the momentum density $\varrho\boldsymbol{p} = \varrho m \boldsymbol{v}$ in equation (\ref{momentum}), where the mass $m$ is constant.  Here $V$ corresponds to the potential surface, where in keeping with the usual formulation in chemical applications in one dimension $V$ may be generally thought of as a model potential (e.g. an Eckart, Lennard-Jones or electrostatic potential). 

The quantum stress $\Pi$ is given to obey, \[\Pi = \varrho m \boldsymbol{v}\otimes\boldsymbol{v} + \varrho^{-1}\bigg\{\frac{\hbar^{2}}{4 m}(\nabla_{x}\varrho)^{2}\bigg\} - \frac{\hbar^{2}}{4m}\nabla_{x}^{2}\varrho,\] or alternatively \[m^{-1} \Pi = \varrho\boldsymbol{v}\otimes\boldsymbol{v} -   \frac{\varrho\hbar^{2}}{4m^{2}}\nabla_{x}^{2}\log\varrho,\] with the Bohmian quantum potential given as $\mathcal{Q} = \left(\frac{\hbar^{2}}{2m}\Delta_{x}\sqrt{\varrho}\right)/\sqrt{\varrho}$  (note that this term is only defined up to a sign convention, see for example Ref.~\cite{Jungel,LM} versus Ref.~\cite{Wyatt}), such that the nonlinear dispersion relation is given by, \begin{equation}\label{bohm} \frac{\varrho\hbar^{2}}{2m}\nabla_{x}\left(\frac{\Delta_{x}\sqrt{\varrho}}{\sqrt{\varrho}}\right)=  \frac{\hbar^{2}}{4m}\nabla\cdot(\varrho\nabla_{x}^{2}\log\varrho), \end{equation} yielding the alternative form of (\ref{momentum}): \begin{equation}\label{momentum2} \partial_{t}(\varrho m\boldsymbol{v}) + \nabla_{x}\cdot(\varrho m\boldsymbol{v}\otimes\boldsymbol{v}) -\varrho\nabla_{x}\mathcal{Q} + \varrho\nabla_{x}V= 0.\end{equation}   

Let us rescale (\ref{mass}) and (\ref{momentum2}) by setting $s=\sqrt{m} t$ and solving for a rescaled solution $\boldsymbol{u}$ and $\rho$ in the time variable $t$, such that $\boldsymbol{u}(t,x) =\sqrt{m}\boldsymbol{v}(\sqrt{m}t,x)$ and $\rho(t,x) = \varrho(\sqrt{m}t,x)$ such that (\ref{mass}) and (\ref{momentum2}) for $(t,x)\in T\times\Omega$ become: \begin{align} \label{massrescale}& \partial_{t} \rho + \nabla_{x}\cdot (\rho \boldsymbol{u}) = 0, \\ \label{momentum2rescale}& \partial_{t}(\rho\boldsymbol{u}) + \nabla_{x}\cdot(\rho\boldsymbol{u}\otimes\boldsymbol{u}) - \rho\nabla_{x}\mathcal{Q} + \rho\nabla_{x}V= 0.\end{align} 

We solve (\ref{massrescale})-(\ref{momentum2rescale}) using a mixed discontinuous Galerkin finite element method.  We define the state vector \[\boldsymbol{U} = (\rho,\rho\boldsymbol{u})^{T},\] the inviscid flux vector \[\boldsymbol{f} =(\rho\boldsymbol{u},\rho\boldsymbol{u}\otimes\boldsymbol{u})^{T},\] and the source vector \[\boldsymbol{S} = \left(0,\rho\nabla_{x}(V -\mathcal{Q})\right)^{T}.\]  Then we can rewrite (\ref{mass})-(\ref{momentum}) as \begin{equation}\label{form1}\boldsymbol{U}_{t} + \boldsymbol{f}_{x} + \boldsymbol{S} = 0.\end{equation} 

\begin{figure}
{\setlength{\unitlength}{4144sp}%
{\begin{picture}(4043,1500)(-800,-400)
{\thinlines
\put(  1,839){\circle*{90}}}%
{\put(  1,839){\circle*{90}}}%
{\put(586,839){\circle*{90}}}%
{\put(1126,839){\circle*{90}}}%
{\put(2296,839){\circle*{90}}}%
{\put(2746,839){\circle*{90}}}%
{\put(3556,839){\circle*{90}}}%
{\put(  1,839){\circle*{90}}}%
{\put(  1,289){\circle{90}}}%
{\put(586,289){\circle{90}}}%
{\put(1126,289){\circle{90}}}%
{\put(2296,289){\circle{90}}}%
{\put(2746,289){\circle{90}}}%
{\put(3556,289){\circle{90}}}%
{\put(586,-151){\circle*{90}}}%
{\put(1126,-151){\circle*{90}}}%
{\put(2296,-151){\circle*{90}}}%
{\put(2746,-151){\circle*{90}}}%
{\put(  1,839){\line( 1, 0){1350}}}%
{\put(3556,839){\line(-1, 0){1485}}}%
{\put( 46,299){\line( 1, 0){495}}}%
{\put(631,299){\line( 1, 0){450}}}%
{\put(1171,299){\line( 1, 0){180}}}%
{\put(2071,299){\line( 1, 0){180}}}%
{\put(2341,299){\line( 1, 0){360}}}%
{\put(2791,299){\line( 1, 0){720}}}%
\put(-44,929){$a$}
\put(3516,929){$b$}
\put(-89,569){$x_0$}
\put(541,569){$x_1$}
\put(1081,569){$x_2$}
\put(1590,279){$\ldots$}
\put(1590,819){$\ldots$}
\put(1590,-171){$\ldots$}
\put(2211,569){$x_{ne-2}$}
\put(2701,569){$x_{ne-1}$}
\put(3501,569){$x_{ne}$}
\put(181,119){$\mathcal{G}_1$}
\put(721,119){$\mathcal{G}_2$}
\put(3016,119){$\mathcal{G}_{ne}$}
\put(2386,119){$\mathcal{G}_{ne-1}$}
\put(451,-421){$\mathcal{K}_{12}$}
\put(991,-421){$\mathcal{K}_{23}$}
\put(2161,-501){$\mathcal{K}_{ne-2,ne-1}$}
\put(2611,-381){$\mathcal{K}_{ne-1,ne}$}
\end{picture}}}
\caption{The discretization of $\Omega$, distinguishing nodes, elements and neighbors, with boundary $\partial\Omega = \{a,b\}$ in dimension $N=1$.}
\label{fig:scheme}
\end{figure}
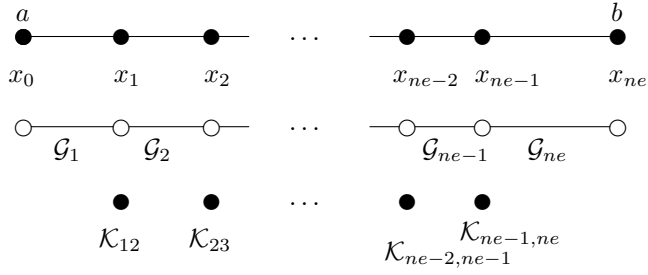

Consider the following discretization scheme motivated by \cite{FFS,MSEV} (and illustrated in the one dimensional case in Figure \ref{fig:scheme}).  Take an open $\Omega\subset\mathbb{R}$ with boundary $\partial\Omega=\Gamma$, given $T>0$ such that $\mathcal{Q}_{T}=((0,T)\times\Omega)$ for $\hat{\Omega}$ the closure of $\Omega$.  Let $\mathscr{T}_{h}$ denote the partition of the closure $\Omega$, such that taking $\hat{\Omega}=[a,b]$ provides the partition \[a=x_{0}<x_{1}\ldots<x_{ne}=b\] comprised of elements $\mathcal{G}_{i}=(x_{i-1},x_{i})\in\mathscr{T}_{h}$ such that $\mathscr{T}_{h}= \{\mathcal{G}_{1},\mathcal{G}_{2}, \ldots,\mathcal{G}_{ne}\}$.  The mesh diameter $h$ is given by $h=\sup_{\mathcal{G}\in\mathscr{T}_{h}}(x_{i}-x_{i-1})$ such that a discrete approximation to $\Omega$ is given by the set $\Omega_{h} = \cup_{i}\mathcal{G}_{i}\setminus\{a,b\}$.  Each element of the partition has a boundary set given by $\partial\mathcal{G}_{i} = \{x_{i-1},x_{i}\}$, where elements sharing a boundary point $\partial\mathcal{G}_{i}\cap\partial\mathcal{G}_{j}\neq\emptyset$ are characterized as neighbors and generate the set $\mathcal{K}_{ij}=\partial\mathcal{G}_{i}\cap\partial\mathcal{G}_{j}$ of interfaces between neighboring elements.  The boundary $\partial\Omega =\{a,b\}$ is characterized in the mesh as $\partial \Omega=\{x_{0},x_{ne}\}$ and indexed by elements $B_{j}\in\partial\Omega$ such that $\hat{\Omega} =\mathscr{T}_{h}\cup\mathcal{K}_{ij}\cup\partial\Omega$.  Now for $I\subset\mathbb{Z}^{+}=\{1,2,\ldots\}$ define the indexing set $r(i)=\{j \in I : \mathcal{G}_{j}$ is a neighbor of $\mathcal{G}_{i}\}$, and for $I_{B}\subset \mathbb{Z}^{-}=\{-1,-2,\ldots\}$ define $s(i)=\{j\in I_{B}:\mathcal{G}_{i}$ contains $B_{j}\}$.  Then for $S_{i}=r(i)\cup s(i)$, we have $\partial\mathcal{G}_{i}=\cup_{j\in S(i)}\mathcal{K}_{ij}$ and $\partial\mathcal{G}_{i}\cap\partial\Omega = \cup_{j\in s(i)}\mathcal{K}_{ij}$.     

We define the broken Sobolev space over the partition $\mathscr{T}_{h}$ as \[W^{k,2}(\Omega_{h},\mathscr{T}_{h})=\{v : v_{|\mathcal{G}_{i}}\in W^{k,2}(\mathcal{G}_{i}) \ \ \forall\mathcal{G}_{i}\in\mathscr{T}_{h}\}.\]  Further, approximate solutions to (\ref{mass})-(\ref{momentum}) will exist in the space of discontinuous piecewise polynomial functions over $\Omega$ restricted to $\mathscr{T}_{h}$, given as \[S_{h}^{d}(\Omega_{h},\mathscr{T}_{h})=\{v:v_{|\mathcal{G}_{i}}\in \mathscr{P}^{d}(\mathcal{G}_{i}) \ \ \forall\mathcal{G}_{i}\in\mathscr{T}_{h}\}\] for $\mathscr{P}^{d}(\mathcal{G}_{i})$ the space of degree $\leq d$ polynomials on $\mathcal{G}_{i}$.

Choosing a set of degree $d$ polynomial basis functions $N_{\ell}\in\mathscr{P}^{d}(\mathcal{G}_{i})$ for $\ell =0,\ldots, d$ we can denote the state vector at the time $t$ over $\Omega_{h}$, by
\begin{equation}\label{shapefunctions}
\boldsymbol{U}_{h}(t,x)=\sum_{\ell=0}^{d}\boldsymbol{U}_{\ell}^{i}(t)N^{i}_{\ell}(x),\quad  \forall x\in\mathcal{G}_{i},
\end{equation}
 where the $N^{i}_{\ell}$'s are the finite element shape functions in the DG setting, and the $\boldsymbol{U}_{\ell}^{i}$'s correspond to the nodal unknowns.   We characterize the finite dimensional test functions \[\boldsymbol{\varphi}_{h}\in W^{2,2}(\Omega_{h},\mathscr{T}_{h}),\quad\mathrm{by}\quad\boldsymbol{\varphi}_{h}(x)=\sum_{\ell=0}^{d}\boldsymbol{\varphi}_{\ell}^{i}N_{\ell}^{i}(x)\] where $\boldsymbol{\varphi}_{\ell}^{i}$ are the nodal values of the test functions in each $\mathcal{G}_{i}$. 

Assuming that the source term $\boldsymbol{S}$ is sufficiently smooth, we let $\boldsymbol{U}$ be a classical solution to (\ref{form1}) and multiply through by $\boldsymbol{\varphi}_{h}$ and integrating such that: \begin{align}\label{solve} & \frac{d}{dt}\int_{\mathcal{G}_{i}}\boldsymbol{U}\cdot\boldsymbol{\varphi}_{h}dx+ \int_{\mathcal{G}_{i}}\boldsymbol{f}_{x}\cdot\boldsymbol{\varphi}_{h}dx = - \int_{\mathcal{G}_{i}}\boldsymbol{S}\cdot\boldsymbol{\varphi}_{h}dx.\end{align}  Integrating (\ref{solve}) by parts gives \begin{equation}\begin{aligned}\label{solveibp}\frac{d}{dt}\int_{\mathcal{G}_{i}}\boldsymbol{U}_{h}\cdot & \boldsymbol{\varphi}_{h}dx+ \int_{\mathcal{G}_{i}}(\boldsymbol{f}\cdot\boldsymbol{\varphi}_{h})_{x}dx-\int_{\mathcal{G}_{i}}\boldsymbol{f}\cdot\boldsymbol{\varphi}^{h}_{x}dx = - \int_{\mathcal{G}_{i}}\boldsymbol{S}\cdot\boldsymbol{\varphi}_{h}dx. \end{aligned}\end{equation} 

Let $\varphi_{|\mathcal{K}_{ij}}$ and  $\varphi_{|\mathcal{K}_{ji}}$ denote the values of $\varphi$ on $\mathcal{K}_{ij}$ considered from the interior and the exterior of $\mathcal{G}_{i}$, respectively.  It should be noted that for $\mathcal{K}_{ij}\in\Gamma$, the restricted functions $\boldsymbol{\varphi}_{h}|\mathcal{K}_{ji}$ are determined up to a choice of boundary condition, which we will discuss in more detail in \textsection{3}.  We approximate the first term in (\ref{solveibp}) by, \begin{equation}
\begin{aligned}\label{timeterm}
\frac{d}{dt}\int_{\mathcal{G}_{i}}\boldsymbol{U}_{h}\cdot \boldsymbol{\varphi}_{h}dx  \approx  \frac{d}{dt}\int_{\mathcal{G}_{i}}\boldsymbol{U}\cdot\boldsymbol{\varphi}_{h}dx,
\end{aligned}
\end{equation} the second term using an inviscid numerical flux $\boldsymbol{\Phi}_{i}$, by
\begin{equation}
\begin{aligned}\label{invflux}
\tilde{\boldsymbol{\Phi}}_{i}(\boldsymbol{U}_{h}|_{\mathcal{K}_{ij}},\boldsymbol{U}_{h}|_{\mathcal{K}_{ji}}, \boldsymbol{\varphi}_{h}) & = \sum_{j\in S(i)}\int_{\mathcal{K}_{ij}}\boldsymbol{\Phi}(\boldsymbol{U}_{h}|_{\mathcal{K}_{ij}},\boldsymbol{U}_{h}|_{\mathcal{K}_{ji}},n_{ij})\cdot\boldsymbol{\varphi}_{h}|_{\mathcal{K}_{ij}} d\mathcal{K} \\ & \approx  \int_{\mathcal{K}_{ij}}\sum_{l =1}^{N}(\boldsymbol{f}_{h})_{l}\cdot (n_{ij})_{l}\boldsymbol{\varphi}_{h}|_{\mathcal{K}_{ij}}d\mathcal{K},
\end{aligned}
\end{equation}
for $n_{ij}$ the unit outward pointing normal and where $l$ is the dimension, and the third term on the left in (\ref{solveibp}) by:
\begin{equation}\label{third}
\boldsymbol{\Theta}_{i}(\boldsymbol{U}_{h},\boldsymbol{\varphi}_{h})= -\int_{\mathcal{G}_{i}} \boldsymbol{f}_{h} \cdot (\boldsymbol{\varphi}_{h})_{x} dx \approx -\int_{\mathcal{G}_{i}} \boldsymbol{f} \cdot (\boldsymbol{\varphi}_{h})_{x} dx.
\end{equation}

Using (\ref{timeterm})-(\ref{third}), taking the convention that \[\mathscr{X} = \sum_{\mathcal{G}_{i}\in\mathscr{T}_{h}}\mathscr{X}_{i},\] and setting the inner product \[(\boldsymbol{a}_{h},\boldsymbol{b}_{h})_{\Omega_{\mathcal{G}}} = \sum_{\mathcal{G}_{i}\in\mathscr{T}_{h}}\int_{\mathcal{G}_{i}}\boldsymbol{a}_{h}\cdot\boldsymbol{b}_{h} dx,\] we define an approximate solution to (\ref{solve})-(\ref{solve3}) as $\boldsymbol{U}_{h}$ for all $t\in (0,T)$ satisfying:

\begin{center} \underline{\emph{Discontinuous Galerkin Method for the QHD Conservation Laws}}\end{center}
\begin{equation}
\begin{aligned}
\label{aprox} 
& 1) \ \boldsymbol{U}_{h}\in C^{0}([0,T]; S_{h}^{d}),\\
& 2) \ \frac{d}{dt}(\boldsymbol{U}_{h},\boldsymbol{\varphi}_{h})_{\Omega_{\mathcal{G}}}+\tilde{\boldsymbol{\Phi}}(\boldsymbol{U}_{h},\boldsymbol{\varphi}_{h})+\boldsymbol{\Theta}(\boldsymbol{U}_{h},\boldsymbol{\varphi}_{h}) + (\boldsymbol{S}_{h},\varphi_{h})_{\Omega_{\mathcal{G}}} = 0, \\ & 3) \ \boldsymbol{U}_{h}(0)=\boldsymbol{U}_{0}.
\end{aligned}
\end{equation}

To compute the source term $\boldsymbol{S}$, we approximate the Bohmian quantum potential using a mixed finite element method.  In particular, we know that at each time $t$, the quantum potential $\mathcal{Q}$ satisfies the equations: \begin{equation}\label{boh}\mathcal{Q}=\frac{\hbar^{2}}{2m}\frac{\nabla_{x}\cdot\boldsymbol{q}}{\sqrt{\rho}} \quad \mathrm{and}\quad \boldsymbol{q} = \nabla_{x}\sqrt{\rho}.\end{equation} Let $\vartheta\in L^{2}(\Omega)$ and $\varsigma\in H(\vdiv,\Omega)$.  Then multiplying (\ref{boh}) by $\vartheta$ and $\varsigma$, respectively, and integrating by parts over $\Omega$ results in: \begin{align} \label{solve2} & \int_{\Omega}\mathcal{Q} \vartheta dx = \int_{\Omega} \frac{\hbar^{2}}{2m}\frac{\nabla_{x}\cdot \boldsymbol{q}}{\sqrt{\rho}} \vartheta dx, \\ \label{solve3} & \int_{\Omega} \boldsymbol{q}\cdot\varsigma dx = -\int_{\Omega}\sqrt{\rho}\nabla_{x}\varsigma dx + \int_{\Gamma}\sqrt{\rho} \varsigma \cdot n d\Gamma.\end{align}

Choosing finite dimensional subspaces $\mathscr{L}^{h}\subset L^{2}(\Omega)$ and $\mathscr{H}^{h}\subset H(\vdiv,\Omega)$, a mixed finite element method for the Bohmian quantum potential is then: find $\mathcal{Q}^{h}:[0,T]\times\Omega\to\mathbb{R},\boldsymbol{q}^{h}:[0,T]\times\Omega\to\mathbb{R}^{3}$ such that for all $t\in[0,T]$,  $\mathcal{Q}^{h}(t)\in\mathscr{L}^{h}$ and  $\boldsymbol{q}^{h}\in \mathscr{H}^{h}$ satisfy:

\begin{center} \underline{\emph{Mixed Method for the Bohmian Quantum Potential}}\end{center}
\begin{equation}
\begin{aligned}
\label{aproxcont} 
& 1) \ (\mathcal{Q}_{h},\vartheta_{h})_{\Omega} = \frac{\hbar^{2}}{m}\left(\frac{\nabla_{x}\cdot\boldsymbol{q}_{h}}{\sqrt{\rho_{h}}},\vartheta_{h}\right)_{\Omega}, \\  & 2) \ (\boldsymbol{q}_{h},\varsigma_{h})_{\Omega} = -(\sqrt{\rho_{h}},\nabla_{x}\varsigma_{h})_{\Omega} + (\sqrt{\rho_{h}},\varsigma_{h}n)_{\Gamma}. 
\end{aligned}
\end{equation}
Since we wish $\boldsymbol{S}\in L^{2}(\Omega)$, we choose $\mathscr{L}^{h}$ to be a continuous finite element space, and we choose $\mathscr{H}^{h}$ to be an $H(\vdiv)$-conforming space (e.g. Raviart-Thomas elements \cite{RT1}, such that in one dimension, Raviart-Thomas elements collapse to be standard continuous finite elements).  Equations (\ref{aprox}) and (\ref{aproxcont}) define our mixed/discontinuous Galerkin method in semi-discrete form.  Computationally, we must also discretize time, as shown in \textsection{4} and \textsection{5}.

It is worth noting that in the Lagrangian formulation the primitive variables $(\rho,\boldsymbol{u})$ are accompanied by the \emph{quantum action} $S$ and the \emph{quantum wave function} $\psi$.  We will explicitly derive these terms in section \textsection{5} from the solution (\ref{aprox}).  It is also worth noting that a pure discontinuous Galerkin method was implemented as an alternative approach to the MDG method solution shown in (\ref{aprox}).  This treatment used a dispersive flux formulation as shown in \cite{LSY}.  We found that this formulation depended nonlinearly on the sign of the advective flux term, leading in the naive implementation to the formation of soliton/compacton type behavior; solutions which are well-known in the `formally' equivalent formulation of Korteweg fluids (see \cite{Kots,DD,Kostin,BDDJ}) -- up to turbulence effects etc., as explained in \textsection{3} -- which model diffuse fluid interfaces as well as having phenomenological interpretation in the context of the nonlinear Schr\"{o}dinger equation (see \cite{TZ}) and the Gross-Pitaevskii equation (see \cite{APA,KV}) given nearly identical initial conditions to the ones we use in \textsection{5}.  However, in the context of chemical dynamics it is not clear that these types of solutions carry physical significance, and so we have isolated our analysis to the MDG method formulation presented in (\ref{aprox}).

\section{\texorpdfstring{\protect\centering $\S 3$ Boundary Treatment}{\S 3 Boundary Treatment}}

A recurring difficulty in constructing numerical methods for initial-boundary value systems of partial differential equations for physical systems is the issue of how to prescribe mathematically consistent boundary conditions which accommodate dynamic (physical) boundary data.  It turns out that this issue is a cause of both numerical and mathematical difficulties in establishing the formal equivalencies between the TDSE and the QHD system of equations.  We show this behavior explicitly in an example in \textsection{5}, but let us first examine the mathematical source of this difficulty.

Recall that the system presented in (\ref{mass})-(\ref{momentum}) is derived explicitly from the TDSE. That is, we have set $\psi = R e^{i S/\hbar}$, and want to expand the solution of the Schr\"{o}dinger equation in one unknown and one equation in $\psi = \psi(t,x)$ into a system of partial differential equations in the unknowns $R=R(t,x)$ and $S=S(t,x)$.  To make this a well-posed system we of course need a system of two equations, where both unknowns must be assigned distinct boundary conditions.  First take the following form of the Schr\"{o}dinger equation: \begin{equation}\label{Sch}\left(-\Delta_{x} + \frac{2m}{\hbar^{2}}V\right)\psi = \frac{2mi}{\hbar}\partial_{t}\psi,\end{equation} and plug in $\psi = R e^{i S/\hbar}$ such that expanding gives for the time derivative, \begin{equation}\begin{aligned} \label{time} \frac{2mi}{\hbar}\partial_{t}\psi  & =  \frac{2mi}{\hbar}\frac{\partial}{\partial t}\left(R e^{iS/\hbar}\right) \\ & =  \frac{2mi}{\hbar} e^{iS/\hbar}\partial_{t}R - \frac{2m}{\hbar^{2}} Re^{iS/\hbar}\partial_{t}S,\end{aligned}\end{equation} and for the spatial component \begin{equation}\begin{aligned}\label{space} \Delta_{x}\psi & = \Delta_{x}(Re^{iS/\hbar}) = \nabla_{x}\cdot\nabla_{x}(Re^{iS/\hbar}) \\ & = \nabla_{x}\cdot\left(e^{iS/\hbar}\nabla_{x}R  + \frac{i}{\hbar} R e^{iS/\hbar}\nabla_{x}S\right) \\ & = e^{iS/\hbar}\left(\Delta_{x}R + \frac{2i}{\hbar}\nabla_{x}S\nabla_{x}R  - \frac{R}{\hbar^{2}}(\nabla_{x}S)^{2} + \frac{i}{\hbar}R\Delta_{x}S \right).\end{aligned}\end{equation} 

Putting (\ref{time}) and (\ref{space}) back into (\ref{Sch}) and canceling a factor of $e^{i S/\hbar}$ we obtain:  \begin{equation}\begin{aligned}\label{Sch2} R\frac{2m}{\hbar^{2}}V = \Delta_{x}R + \frac{2i}{\hbar}\nabla_{x}S\nabla_{x}R  - \frac{R}{\hbar^{2}}(\nabla_{x}S)^{2} + \frac{i}{\hbar}R\Delta_{x}S + \frac{2mi}{\hbar}\partial_{t}R - \frac{2m}{\hbar^{2}} R \partial_{t}S,\end{aligned}\end{equation}  Now, collecting the imaginary parts of (\ref{Sch2}), \[-\frac{2mi}{\hbar}\partial_{t}R - \frac{2i}{\hbar}\nabla_{x}S\nabla_{x}R - \frac{i}{\hbar}R\Delta_{x}S =0,\] and multiplying through by $\hbar^{2}/2m$ provides: \[-\partial_{t}R -  \frac{1}{m}\nabla_{x}R\nabla_{x}S - \frac{1}{2m}R\Delta_{x}S = 0.\]  Additionally multiplying through by $-2mR$ gives, \[m\partial_{t}R^{2} + \nabla_{x}R^{2}\nabla_{x}S + R^{2}\Delta_{x}S = 0,\] where applying the product rule yields the conservation form:  \begin{equation} m\partial_{t}R^{2} + \nabla_{x}\cdot(R^{2}\nabla_{x}S) = 0.\end{equation}  Clearly setting $R=\sqrt{\varrho}$ and using the Madelung relation $\boldsymbol{v} = \frac{1}{m}\nabla_{x}S$ for $m$ a constant $m\in\mathbb{R}$ leads to the usual conservation of mass equation: \begin{equation}\label{classmass}\partial_{t}\varrho + \nabla_{x}\cdot(\varrho\boldsymbol{v}) = 0.\end{equation}

Similarly putting together the real parts of (\ref{Sch2}) gives: \[ \frac{2mR}{\hbar^{2}}\partial_{t}S  - \Delta_{x}R + \frac{R}{\hbar^{2}}(\nabla_{x}S)^{2} + \frac{2m}{\hbar^{2}}RV = 0,\] such that upon multiplication through by $\hbar^{2}/2 m^{2}R$ we have:\[ \frac{1}{m}\partial_{t}S  - \frac{\hbar^{2}}{2m^{2}R}\Delta_{x}R + \frac{1}{2m^{2}}(\nabla_{x}S)^{2} + \frac{1}{m}V= 0.\] Taking a derivation in $x$ then yields \[ \frac{1}{m}\partial_{t}\nabla_{x}S  - \nabla_{x}\left(\frac{\hbar^{2}}{2m^{2}R}\Delta_{x}R\right) + \nabla_{x}\cdot\left(\frac{1}{2m^{2}}(\nabla_{x}S)^{2}\right) + \frac{1}{m}\nabla_{x}V= 0.\] Now again we substitute the important Madelung relation $\boldsymbol{v} = \frac{1}{m}\nabla_{x}S$ giving the form:  \begin{equation} \label{irro}\partial_{t}\boldsymbol{v}  +\frac{1}{2} \nabla_{x}(\boldsymbol{v}\cdot\boldsymbol{v}) - \frac{\hbar^{2}}{2m^{2}}\nabla_{x}(R^{-1}\Delta_{x}R)  + \frac{1}{m}\nabla_{x}V= 0.\end{equation}  The Madelung relation, $\boldsymbol{v} = \frac{1}{m}\nabla_{x}S$, is of course equivalent to setting $\boldsymbol{v}$ to be an irrotational field, since for \emph{any} field $S$, $\nabla_{x}\times\nabla_{x} S =0.$  Thus for an irrotational vector field $\boldsymbol{v}$, using that $\nabla_{x}(\boldsymbol{v}\cdot\boldsymbol{v}) = 2((\boldsymbol{v}\cdot\nabla_{x})\boldsymbol{v} + \boldsymbol{v}\times\nabla_{x}\times\boldsymbol{v})$, we may rewrite (\ref{irro}) as, \[ \partial_{t}\boldsymbol{v}  + (\boldsymbol{v} \cdot \nabla_{x})\boldsymbol{v} - \frac{\hbar^{2}}{2m^{2}}\nabla_{x}(R^{-1}\Delta_{x}R)  + \frac{1}{m}\nabla_{x}V= 0,\] so that multiplying by $\varrho m$ yields, \[ \varrho \partial_{t}m\boldsymbol{v}  + (\varrho m\boldsymbol{v}\cdot \nabla_{x}) \boldsymbol{v} - \varrho \frac{\hbar^{2}}{2m}\nabla_{x}(R^{-1}\Delta_{x}R)  + \varrho \nabla_{x}V= 0.\]  Combining this equation with (\ref{classmass}) yields: \begin{equation}\label{momentum0} \partial_{t}(\varrho m\boldsymbol{v}) + \nabla_{x}\cdot(\varrho m\boldsymbol{v}\otimes\boldsymbol{v}) -\varrho\nabla_{x}\mathcal{Q} + \varrho\nabla_{x}V= 0,\end{equation} for $\mathcal{Q}$ the Bohmian quantum potential given as $\mathcal{Q} = \left(\frac{\hbar^{2}}{2m}\Delta_{x}\sqrt{\varrho}\right)/\sqrt{\varrho}$.  It is important to see that the formal equivalence between \ref{Sch} and \ref{momentum0} is entirely dependent on Madelung's irrotational condition, which makes turbulent effects, for example, vanish.  In the alternative derivation of the QHD regime, using moment expansions (see for example \cite{Jungel,Wyatt}) this restriction is not necessary.

Thus we have arrived at our system of quantum hydrodynamic equations: \begin{equation}\begin{aligned}\label{QHDnew}& \partial_{t}\varrho + \nabla_{x}\cdot(\varrho\boldsymbol{v}) = 0, \\ &  \partial_{t}(\varrho m\boldsymbol{v}) + \nabla_{x}\cdot(\varrho m\boldsymbol{v}\otimes\boldsymbol{v}) -\varrho\nabla_{x}\mathcal{Q} + \varrho\nabla_{x}V= 0,\end{aligned}\end{equation} requiring initial conditions \[\rho_{|t=0}=\rho_{0}\quad\mathrm{and}\quad u_{|t=0}=u_{0},\] and numerically requiring explicit boundary conditions $\rho_{b}$ and $u_{b}$ on an irrotational vector field $\boldsymbol{v}$.  Additionally, and as an important aside, the formal equivalence we have derived is constructed without mention of boundary conditions, which is satisfied over $(0,T)\times\mathbb{R}^{3}$, but on a discrete domain $\Omega\subset\mathbb{R}^{3}$ is a bit over optimistic, and as we will see below, does not in general hold.  

That is, the TDSE code (see \textsection{5}) sets the initial data $\psi_{i,b}$ on the boundary as a time-independent condition, so the boundary value $\psi_{b}\equiv\psi_{b}=\psi_{i,b}$ is enforced for all $t\in [0,T)$.  Since $\psi_{i,b}$ must be decomposed into $R_{i,b}$ and $S_{i,b}$ to make sense for the QHD formulation (\ref{QHDnew}), these give Dirichlet conditions which can be implemented, but are unstable in the QHD regime, since $R_{i,b}$ exponentially decays on the boundary and as a consequence is not numerically invertible; as it must be in the QHD formulation.  These may however be approximated by setting $\rho_{i,b}=\rho_{A}$, the ambient density, and $u_{i,b}=-\frac{1}{m}\int_{\mathcal{G}_{b}}\nabla S dx$ for $\mathcal{G}_{b}$ the boundary element.

However, these BCs still are not well-posed in the QHD regime for the following reason.  First we compute the entropy inequality for the rescaled version of (\ref{QHDnew}) shown in (\ref{massrescale}) and (\ref{momentum2rescale}).  We may compute the important classical/quantum entropy satisfying for non-boundary terms that: \begin{equation}\label{nonboundentropy}\frac{d}{dt}\int_{\Omega}\left(\rho\frac{|\boldsymbol{v}|^{2}}{2} + \frac{\hbar^{2}(\nabla_{x}\sqrt{\rho})^{2}}{4m} + \rho V\right) dx \leq 0. \end{equation}  

We arrive at this system by multiplying the momentum equation from (\ref{QHDnew}) by $\boldsymbol{v}$ and integrating in space (e.g. the domain is some $\Omega \subseteq \mathbb{R}^{3}$), such that rearranging we find \begin{equation}\label{classicalquantum} \int_{\Omega}\boldsymbol{v}\partial_{t}(\rho\boldsymbol{v}) + \boldsymbol{v}\nabla_{x}\cdot (\rho\boldsymbol{v}\otimes\boldsymbol{v} )dx -\int_{\Omega}\rho\boldsymbol{v}\nabla_{x}\mathcal{Q} dx + \int_{\Omega}\rho\boldsymbol{v}\nabla_{x}V dx = 0.\end{equation}  The product rule allows us to expand the first term on the LHS as: \[\int_{\Omega}|\boldsymbol{v}|^{2}(\partial_{t}\rho + \nabla_{x}\cdot(\rho\boldsymbol{v})) + \rho\boldsymbol{v}\partial_{t}\boldsymbol{v}  + \rho|\boldsymbol{v}|^{2}\nabla_{x}\cdot \boldsymbol{v} \ dx,  \] where $|\boldsymbol{v}|^{2} = \boldsymbol{v}\cdot\boldsymbol{v}$.  Using the mass conservation equation twice from (\ref{QHDnew}) and applying the divergence theorem we find that, \begin{equation}\label{firstone}\int_{\Omega}\boldsymbol{v}\left(\partial_{t}(\rho\boldsymbol{v}) + \nabla\cdot (\rho\boldsymbol{v}\otimes\boldsymbol{v} )\right) dx = \frac{d}{dt}\int_{\Omega}\rho\frac{|\boldsymbol{v}|^{2}}{2}dx + \frac{1}{2}\int_{\Omega}\nabla_{x}\cdot(\rho\boldsymbol{v}^{3})dx.\end{equation}  Next, using the dispersion relation from the Bohm quantum potential the third term on the left yields: \begin{equation}\begin{aligned}\label{secondone}\int_{\Omega}\rho\boldsymbol{v}\nabla_{x}\mathcal{Q} & =  \frac{\hbar^{2}}{2m}\int_{\Omega}\rho\boldsymbol{v}\nabla_{x}\left(\frac{\Delta_{x}\sqrt{\rho}}{\sqrt{\rho}}\right) dx  \\ & =   - \frac{\hbar^{2}}{2m}\int_{\Omega}\frac{1}{\sqrt{\rho}} \nabla_{x}\cdot(\rho\boldsymbol{v}) \Delta_{x}\sqrt{\rho} dx  + \frac{\hbar^{2}}{2m}\int_{\Omega}\nabla_{x}\cdot\left(\sqrt{\rho}\boldsymbol{v}\Delta_{x}\sqrt{\rho}\right) dx \\  & =  \frac{\hbar^{2}}{2m}\int_{\Omega}\bigg\{\nabla_{x}\left(\frac{1}{\sqrt{\rho}} \nabla_{x}\cdot(\rho\boldsymbol{v})\right) \nabla_{x}\sqrt{\rho} + \nabla_{x}\cdot\left(\sqrt{\rho}\boldsymbol{v}\Delta_{x}\sqrt{\rho}\right)\bigg\}dx \\ &\quad\quad  - \frac{\hbar^{2}}{2m}\int_{\Omega}\nabla_{x}\left(\frac{1}{\sqrt{\rho}} \nabla_{x}\cdot(\rho\boldsymbol{v}) \nabla_{x}\sqrt{\rho}\right) dx \\ & = \frac{\hbar^{2}}{2m}\int_{\Omega}\nabla_{x}\sqrt{\rho} \ \partial_{t}\nabla_{x}\sqrt{\rho}dx + \mathrm{boundary} \ \mathrm{terms}.\end{aligned}\end{equation}  Finally the source term $V=V(x)$ upon integrating by parts gives \begin{equation}\begin{aligned}\label{thirdone}\int_{\Omega}\rho\boldsymbol{v}\cdot\nabla_{x} V dx & = -\int_{\Omega}V\nabla_{x}\cdot(\rho\boldsymbol{v})dx + \int_{\Omega}\nabla_{x}\cdot(V\rho\boldsymbol{u})dx \\ & =  \frac{d}{dt}\int_{\Omega}\rho V dx + \int_{\Omega}\nabla_{x}\cdot(V\rho\boldsymbol{v})dx.\end{aligned}\end{equation}

Then we have recovered (\ref{nonboundentropy}) as an equality up to the boundary terms in (\ref{firstone}), (\ref{secondone}) and (\ref{thirdone}).  To recover the mathematical well-posedness of the system these boundary terms must either vanish or be bounded and positive (or negative) definite.  One such choice of boundary data is, for example, $\boldsymbol{v}_{b}\equiv 0$.  Another is the pair of conditions $\nabla_{x}\sqrt{\rho_{b}}\equiv 0 $ and $V_{b}\equiv 0$ for all $t\in[0,T)$, and so forth.  

The first set of boundary data, with $\boldsymbol{v}_{b}\equiv 0$, may be set with $\rho_{b}\equiv\rho_{A}$.  Since the \emph{action} behaves as a phase, this seems a reasonable approximation, since it effectively assumes that up to a constant of integration that the phase is constant over boundary elements $\nabla S_{b}\equiv 0$.  These conditions are then mathematically consistent with the system of equations (\ref{QHDnew}), but have the physical effect of generating ``\emph{inlet/outlet}'' boundary layers, caused by the value of $\rho_{b}$.  

Perhaps a more natural boundary condition is given by setting, \[\boldsymbol{U}_{h}^{n}|_{\mathcal{K}_{ji}}=\boldsymbol{U}_{h}^{n}|_{\mathcal{K}_{ij}},\] where $\boldsymbol{U}_{h}^{n}$ is the numerical solution at timestep $t_{n}$, as explained in detail in \textsection{4}, and $\mathcal{K}_{ij}\in\partial\Omega$.  This boundary type is a first order approximation to a transmissive or radiative condition that treats the boundary like a ``ghost cell,'' and allows density and momentum to leave the domain as though falling into vacuum, while allowing no density or momentum to enter.  This condition approximates to the first order, the effect of ``not setting boundary conditions at all,'' and thus \emph{not badly} perturbing the system (\ref{QHDnew}) away from its natural behavior, nor generating reflecting behavior, which in some contexts -- such as a chemical reaction occurring in a solvent bath -- are difficult to physically interpret. 

\section{\texorpdfstring{\protect\centering $\S 4$ A Numerical Test Case}{\S 4 A Numerical Test Case}}

We wish to test the accuracy of our MDG method formulation by solving an analytic test solution.  In order to do this we choose a numerical flux for (\ref{aprox}) and restrict to spatial dimension $l=1$. For the inviscid flux $\boldsymbol{\Phi}$ we implement the local Lax-Friedrich's flux $\boldsymbol{\Phi}_{lLF}$ satisfying
\[
\begin{aligned}
\int_{\mathcal{K}_{ij}}\boldsymbol{\Phi}_{lLF}\cdot\boldsymbol{\varphi}_{h} d\mathcal{K} & =  \frac{1}{2}\int_{\mathcal{K}_{ij}}(\boldsymbol{f}(\boldsymbol{U}_{h})_{|\mathcal{K}_{ij}}+\boldsymbol{f}(\boldsymbol{U}_{h})_{|\mathcal{K}_{ji}})\cdot n_{ij}\boldsymbol{\varphi}_{h}|_{\mathcal{K}_{ij}}d\mathcal{K} \\
& - \frac{1}{2}\int_{\mathcal{K}_{ij}}(\Spec_{r}(\boldsymbol{\Gamma}_{0}))((\boldsymbol{U}_{h})_{|\mathcal{K}_{ij}}-(\boldsymbol{U}_{h})_{|\mathcal{K}_{ji}})\cdot n_{ij}\boldsymbol{\varphi}_{h}|_{\mathcal{K}_{ij}} d\mathcal{K},
\end{aligned}
\]
for $n_{ij}$ the outward unit normal and $\Spec_{r}(\boldsymbol{\Gamma}_{0})$ the spectral radius of $\boldsymbol{\Gamma}_{0}$; the Jacobian matrix of the inviscid flux $J_{\boldsymbol{U}}\boldsymbol{f}(\boldsymbol{U})=\boldsymbol{\Gamma}_{0}(\boldsymbol{U})$ which may be represented by the following $2\times 2$ matrix,
\begin{equation}
\label{jac}
\boldsymbol{\Gamma}_{0}(\boldsymbol{U}) = \left( \begin{array}{cc}
   0    & 1  \\
   -u^2 & 2u \\
   \end{array} \right).
\end{equation}
 Summing over the elements of the mesh this term satisfies:
\begin{equation}
\begin{aligned}
\label{dislax}
2\tilde{\boldsymbol{\Phi}}_{lLF}(\boldsymbol{U}_{h}, \boldsymbol{\varphi}_{h}) = & \sum_{\mathcal{G}_{i}\in\mathscr{T}_{h}}\sum_{j\in S(i)}\int_{\mathcal{K}_{ij}}(\boldsymbol{f}(\boldsymbol{U}_{h})_{|\mathcal{K}_{ij}}+\boldsymbol{f}(\boldsymbol{U}_{h})_{|\mathcal{K}_{ji}})\cdot n_{ij}\boldsymbol{\varphi}_{h}|_{\mathcal{K}_{ij}}d\mathcal{K} \\ - \sum_{\mathcal{G}_{i}\in\mathscr{T}_{h}}&\sum_{j\in S(i)} \int_{\mathcal{K}_{ij}}(\Spec_{r}(\boldsymbol{\Gamma}_{0}))((\boldsymbol{U}_{h})_{|\mathcal{K}_{ij}}-(\boldsymbol{U}_{h})_{|\mathcal{K}_{ji}})\cdot n_{ij}\boldsymbol{\varphi}_{h}|_{\mathcal{K}_{ij}} d\mathcal{K}.
\end{aligned}
\end{equation}

\begin{figure}[!t]
\centering
\includegraphics[width=6cm]{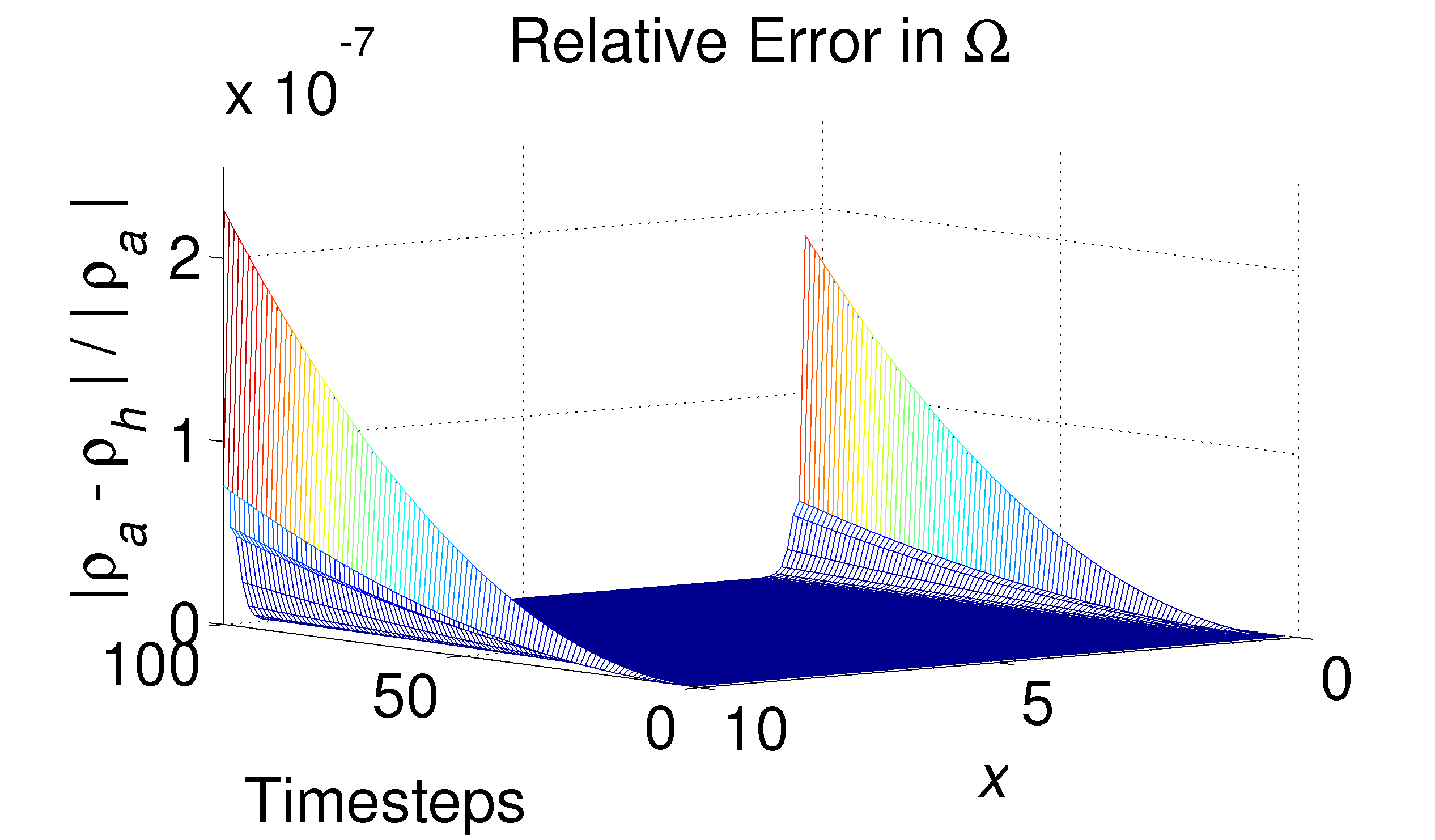} \includegraphics[width=6cm]{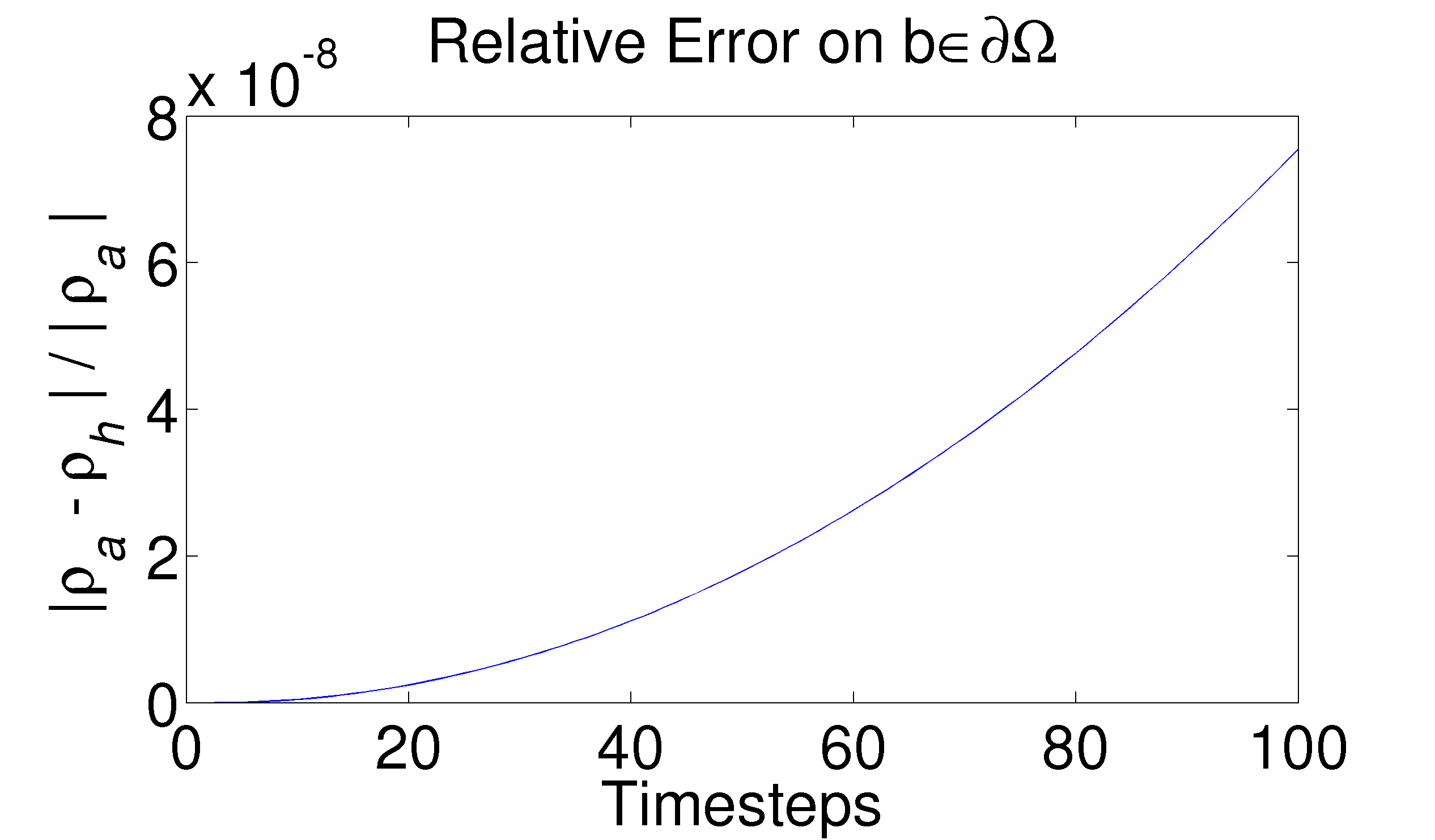}\\ \includegraphics[width=6cm]{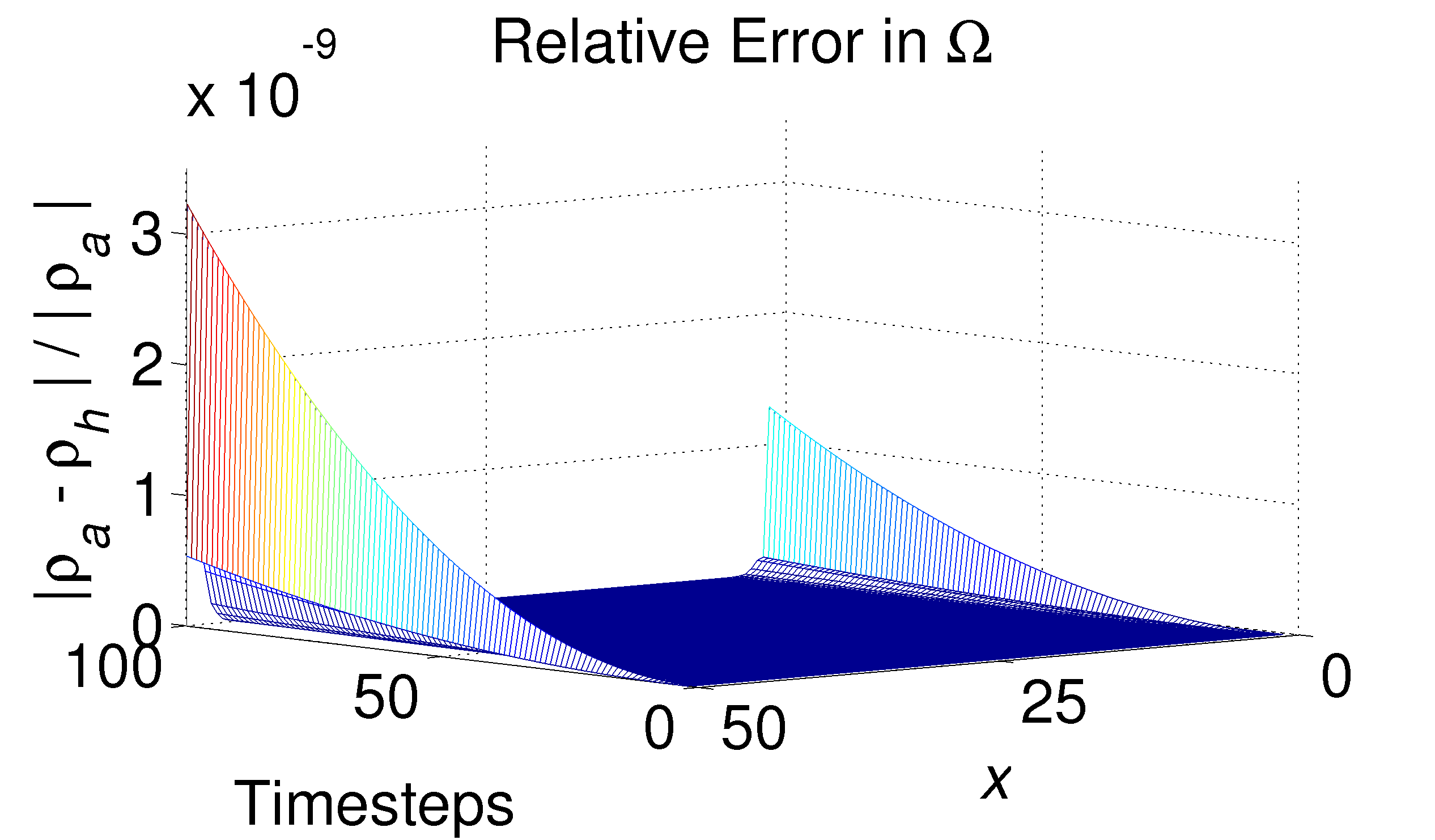} \includegraphics[width=6cm]{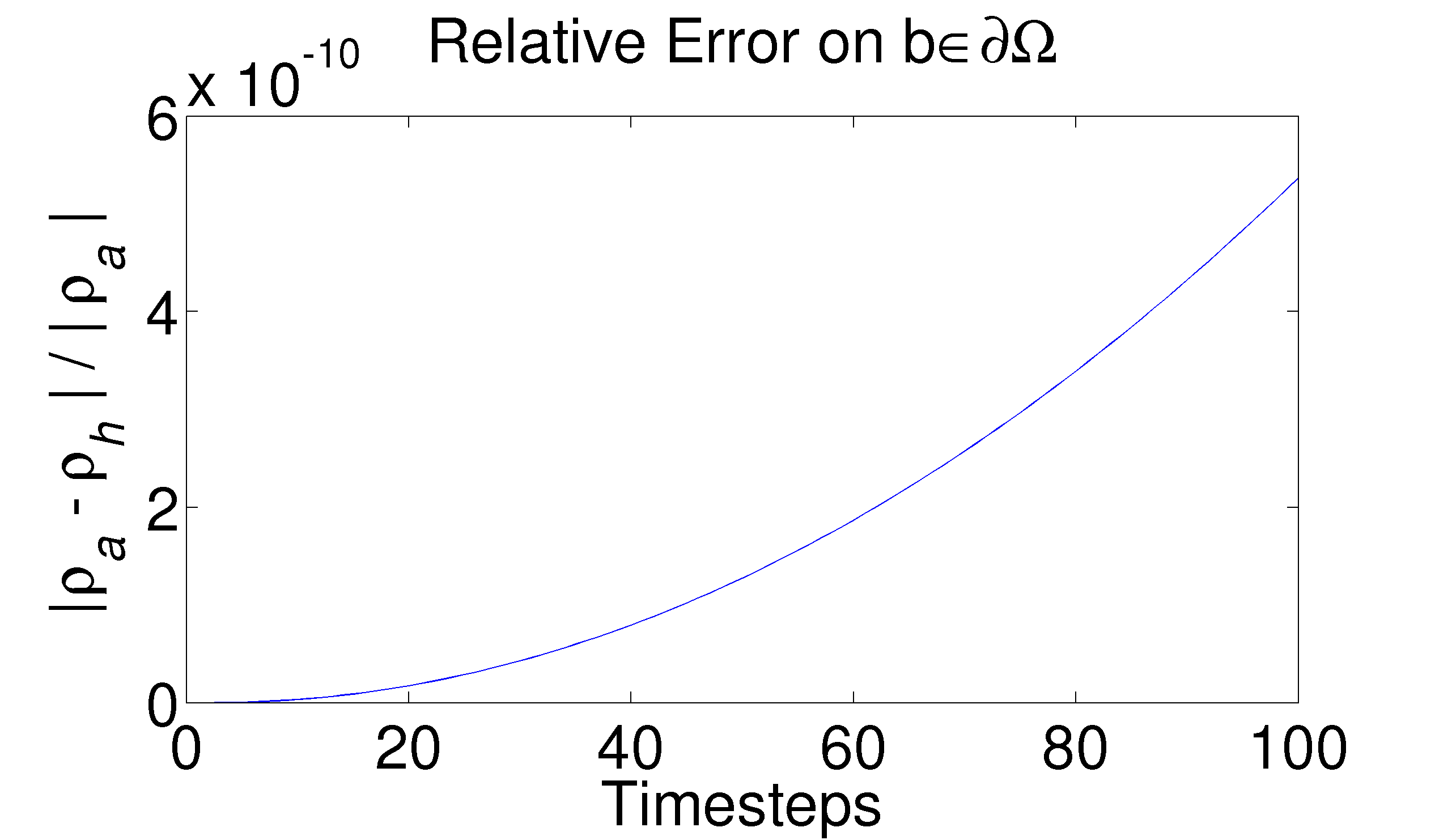}
\caption{Here we show the relative error introduced by the \emph{weak entropy} boundary conditions for $a=0$ and both $b=10$ and $b=50$. The boundary data (the graphs on the right) show only the relative error on element $b$ of $\partial\Omega$ for $b=10$ and $b=50$, respectively.}
\label{fig:test}  
\end{figure}

Next we discretize in time.  That is, we denote a partition of [0,T] by \[0=t^{0}<t^{1}\ldots<t^{T}=T,\] for a timestep given as $\Delta t^{n}=t^{n+1}-t^{n}$., and let $\boldsymbol{U}_{h}^{n}$ denote the solution at timestep $t^{n}$.  Thus we implement the following forward Euler scheme:
\[
\frac{\partial\boldsymbol{U}_{h}}{\partial t} \approx \frac{\boldsymbol{U}_{h}^{n+1}-\boldsymbol{U}_{h}^{n}}{\Delta t^{n}}, 
\] which, along with the implementation of a slope limiter in the conservation variables $(\rho, \rho u)$ given by van Leer's MUSCL scheme (as shown in \cite{VL1} and \cite{VL2}), allows us to explicitly solve (\ref{aprox}).  That is, we define an approximate solution as $\boldsymbol{U}_{h}^{n}$ for all $t_{n}$ such that $n=0,\ldots, T$ satisfying: \begin{equation}
\begin{aligned}
\label{aprox3}
& 1) \ \boldsymbol{U}^{n}_{h}\in S_{h}^{d}, \ \mathcal{Q}_{h}^{n}\in\mathscr{L}_{h} \ \mathrm{and} \ \boldsymbol{q}_{h}^{n}\in\mathscr{H}_{h}, \\
& 2) \ \left(\frac{\boldsymbol{U}_{h}^{n+1}- \boldsymbol{U}_{h}^{n}}{\Delta t^{n}}, \boldsymbol{\varphi}_{h}\right)_{\Omega_{\mathcal{G}}} +\tilde{\boldsymbol{\Phi}}(\boldsymbol{U}_{h}^{n},\boldsymbol{\varphi}_{h})+\boldsymbol{\Theta}(\boldsymbol{U}_{h}^{n},\boldsymbol{\varphi}_{h}) + (\boldsymbol{S}_{h}^{n},\varphi_{h})_{\Omega_{\mathcal{G}}} = 0, \\  & 3) \ (\mathcal{Q}_{h}^{n},\vartheta_{h})_{\Omega} = \frac{\hbar^{2}}{m}\left(\frac{\nabla_{x}\cdot\boldsymbol{q}_{h}^{n}}{\sqrt{\rho^{n}_{h}}},\vartheta_{h}\right)_{\Omega}, \\  & 4) \ \left(\boldsymbol{q}_{h}^{n},\varsigma_{h}\right)_{\Omega} = -\left(\sqrt{\rho_{h}^{n}},\nabla_{x}\varsigma_{h}\right)_{\Omega} + \left(\sqrt{\rho_{h}^{n}},\varsigma_{h}n\right)_{\Gamma}, \\
& 5) \ \boldsymbol{U}_{0}^h=\boldsymbol{U}_{h}(0).
\end{aligned}
\end{equation}  The above formulation lends itself naturally to a staggered scheme.  First, given $\boldsymbol{U}_{h}^{n}$ one solves step 3 and 4 for $\mathcal{Q}_{h}^{n}$ and $\boldsymbol{q}_{h}^{n}$, which provides $\boldsymbol{S}_{h}^{n}$, allowing us to solve for $\boldsymbol{U}_{h}^{n+1}$ in step 2.

Now we construct an appropriate test case.  Consider the dimension $N=1$ case and let $u\equiv 0$ on $\Omega$ for  (\ref{massrescale})-(\ref{momentum2rescale}), such that $\partial_{s}\rho = 0$.  Up to a choice of boundary conditions, upon integration we have for (\ref{momentum2rescale}) that \[\mathcal{Q}=C-V,\] such that choosing a $C\equiv V$ we find the following second order ordinary differential equation: \[\rho'' - \rho^{-1}(\rho')^{2} = 0,\] whose solution is $\rho =e^{x}$.  We solve for the approximate solution of (\ref{aprox}) using the above scheme, with initial conditions $\rho_{0}=e^{x},u=0$, $V=C$ and $m=1836$ the mass of a proton in Hartree atomic units (au).  The boundaries are set to the \emph{weak entropy} boundary condition formulation as presented in \cite{Cockburn1,BLN,Martin} and \cite{MSEV}.  We graph the relative error of our approximate solution $\rho_{h}$ to the exact numerical representation $\rho_{a}$ in Figure (\ref{fig:test}).  We see that the two solutions are numerically exact in the interior of the domain, and error accumulates in the boundary $\partial\Omega$, as expected due to the weak entropy boundary conditions.  We note that the error on the boundary may be reduced by increasing the absolute size of the interval $[a,b]$.

\section{\texorpdfstring{\protect\centering $\S 5$ Tunneling in TDSE and QHD}{\S 5  Tunneling in TDSE and QHD}}

We proceed by testing a relatively standard example in quantum chemistry, given by a propagating Gaussian packet in the direction of a model Eckart potential barrier.  We solve the following one dimensional system:  \begin{align} \label{massrescale1} & \partial_{t} \rho + \partial_{x}(\rho u) = 0, \\ \label{momentum2rescale1}& \partial_{t}(\rho u) + \partial_{x}(\rho u^{2}) - \rho\partial_{x}\mathcal{Q} + \rho\partial_{x}V= 0.\end{align} with initial conditions \begin{equation}\label{init1}\rho_{0}=\rho_{A}+\left(\frac{1}{\sqrt{2\pi\mu}}\right)e^{\frac{-(x-x_{0})^{2}}{2\mu}}\quad\mathrm{and}\quad u_{0} = \left(\alpha V_{0}\right)^{1/2},\end{equation} where the Eckart potential is given by \begin{equation}\label{init2}V(x)=V_{0} \ \sech^{2}\left(\frac{1}{2}(x-x_{1})\right).\end{equation} 

\begin{figure}[!t]
\centering
\includegraphics[width=6cm]{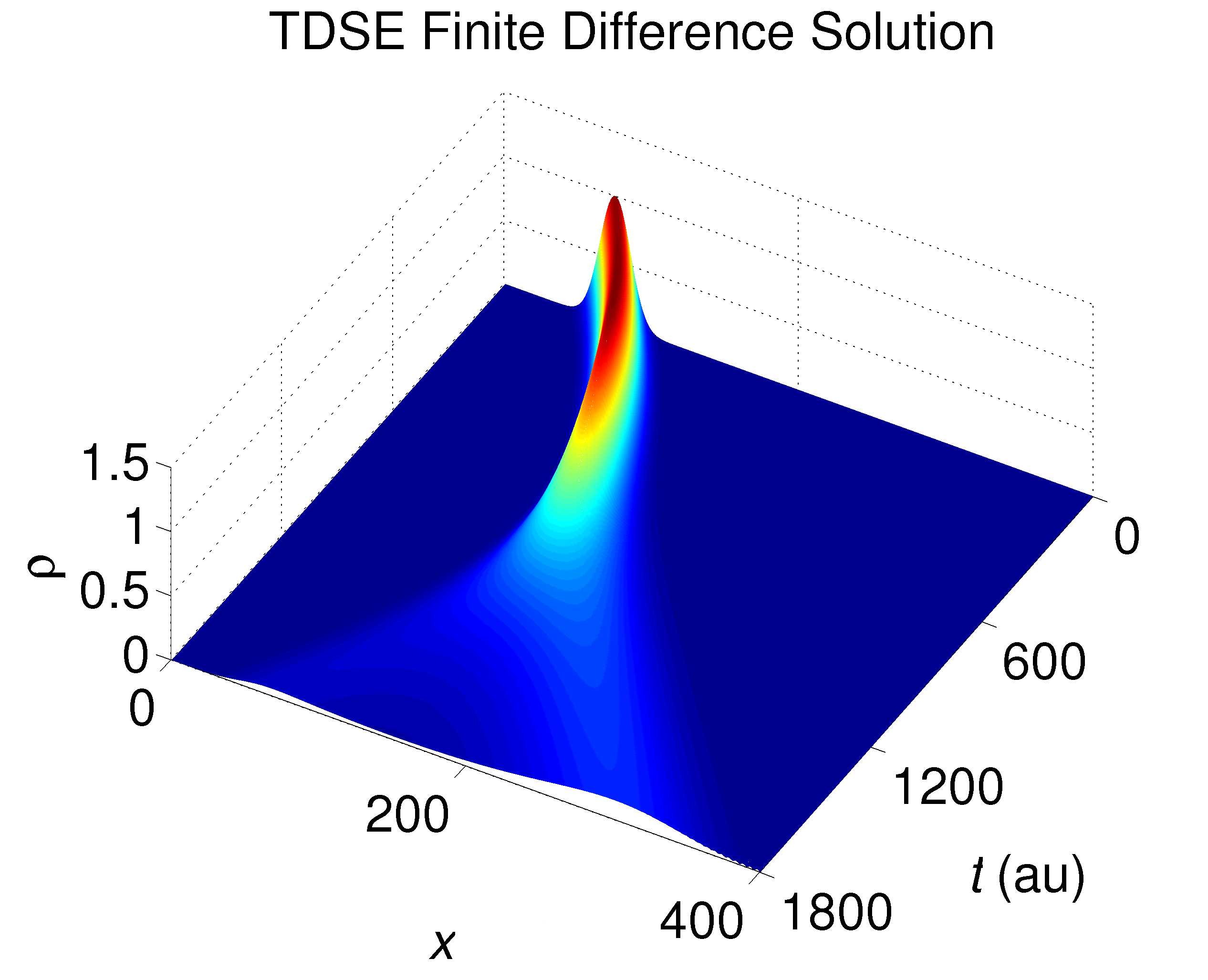}\includegraphics[width=6cm]{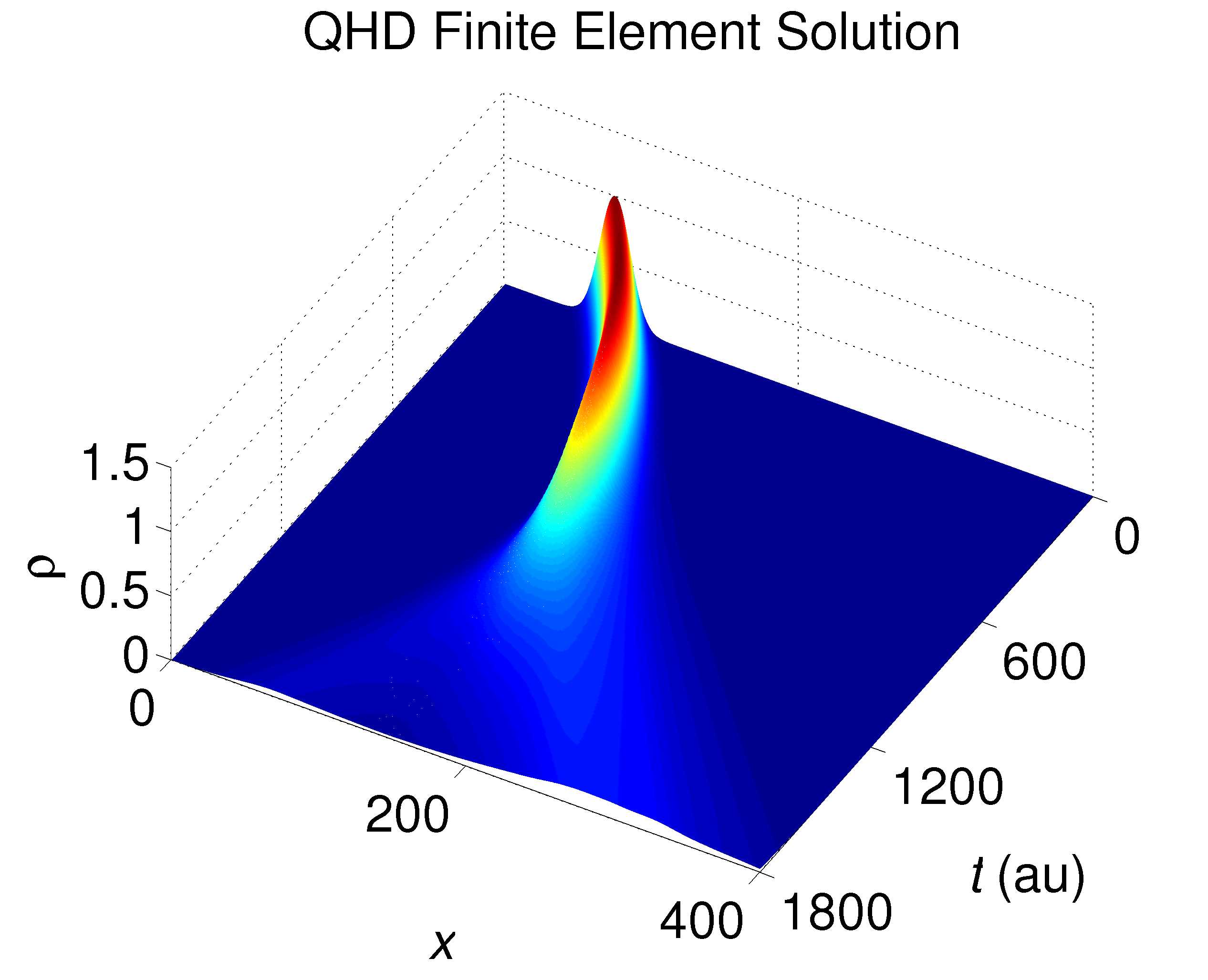}\\ \includegraphics[width=10cm]{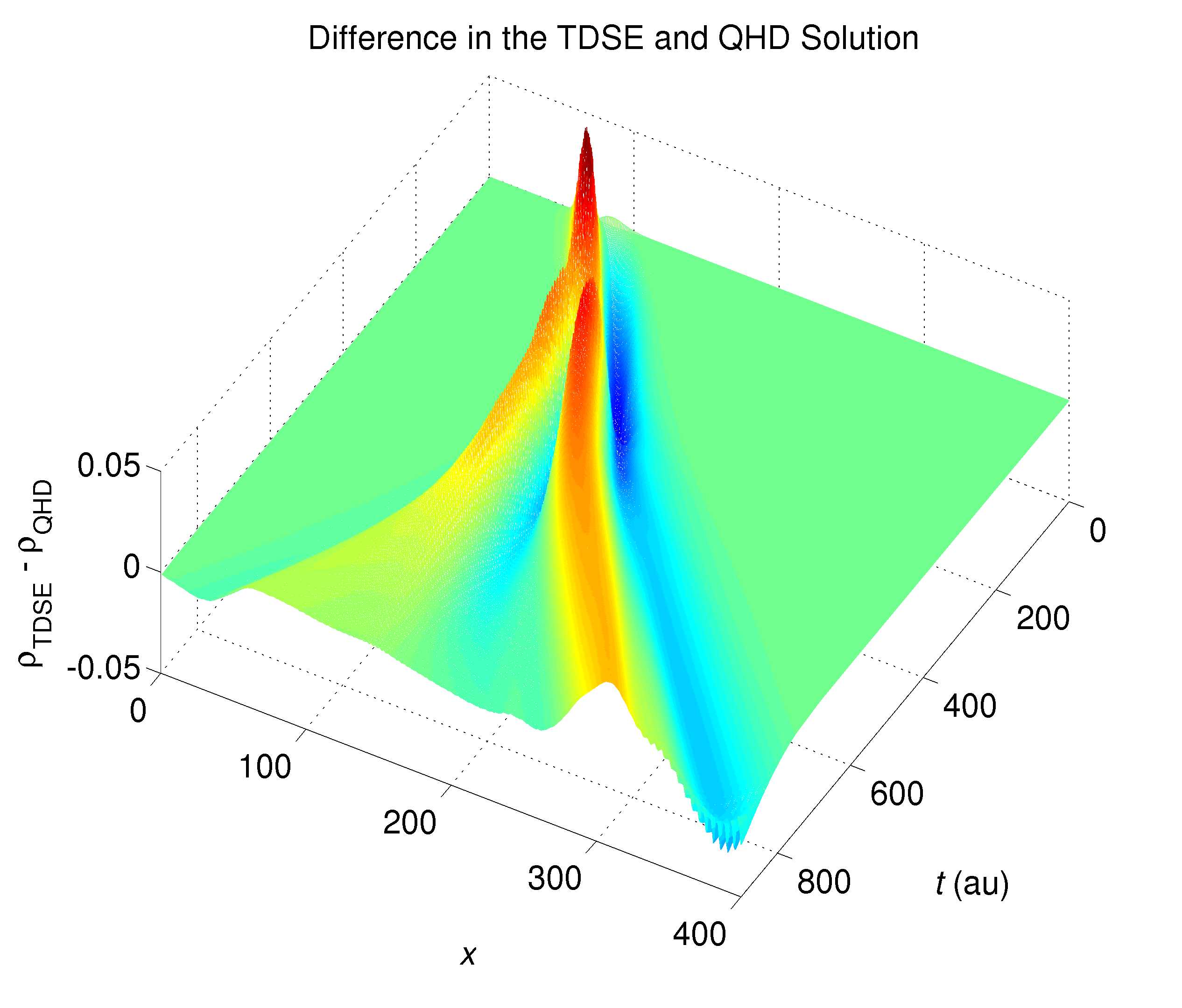} \caption{The top graphs compare solutions to the TDSE and QHD system in the so-called ``eyeball norm,'' for the forward Euler scheme.  The bottom solution shows the nontrivial formal difference.  Here $x$ refers to the $x$-th meshpoint.}
\label{fig:class}  
\end{figure}

As is conventional in quantum hydrodynamics, the mass is set to approximate the hydrogen (proton) mass $m\sim 2000$ au (in Hartree atomic units), $\rho_{A}\sim 10^{-10}$ is a numerical background density for division, $x_{0}$ centers the Gaussian packet, $x_{1}$ centers the potential, $\mu$ is the variance of the distribution, $\alpha$ is a constant $\alpha\in\mathbb{R}$ and $V_{0}$ is the barrier height (which we may vary, so some constant $V_{0}\in\mathbb{R}$).  In the quantum regime (when classical barrier transmission is not present), the initial velocity $u_{0}$ is often chosen to satisfy the following condition on the initial kinetic energy $K_{0}=\frac{1}{2}u_{0}^{2}=\frac{1}{4}V_{0}$.

\begin{figure}[!t]
\centering
\includegraphics[width=6cm]{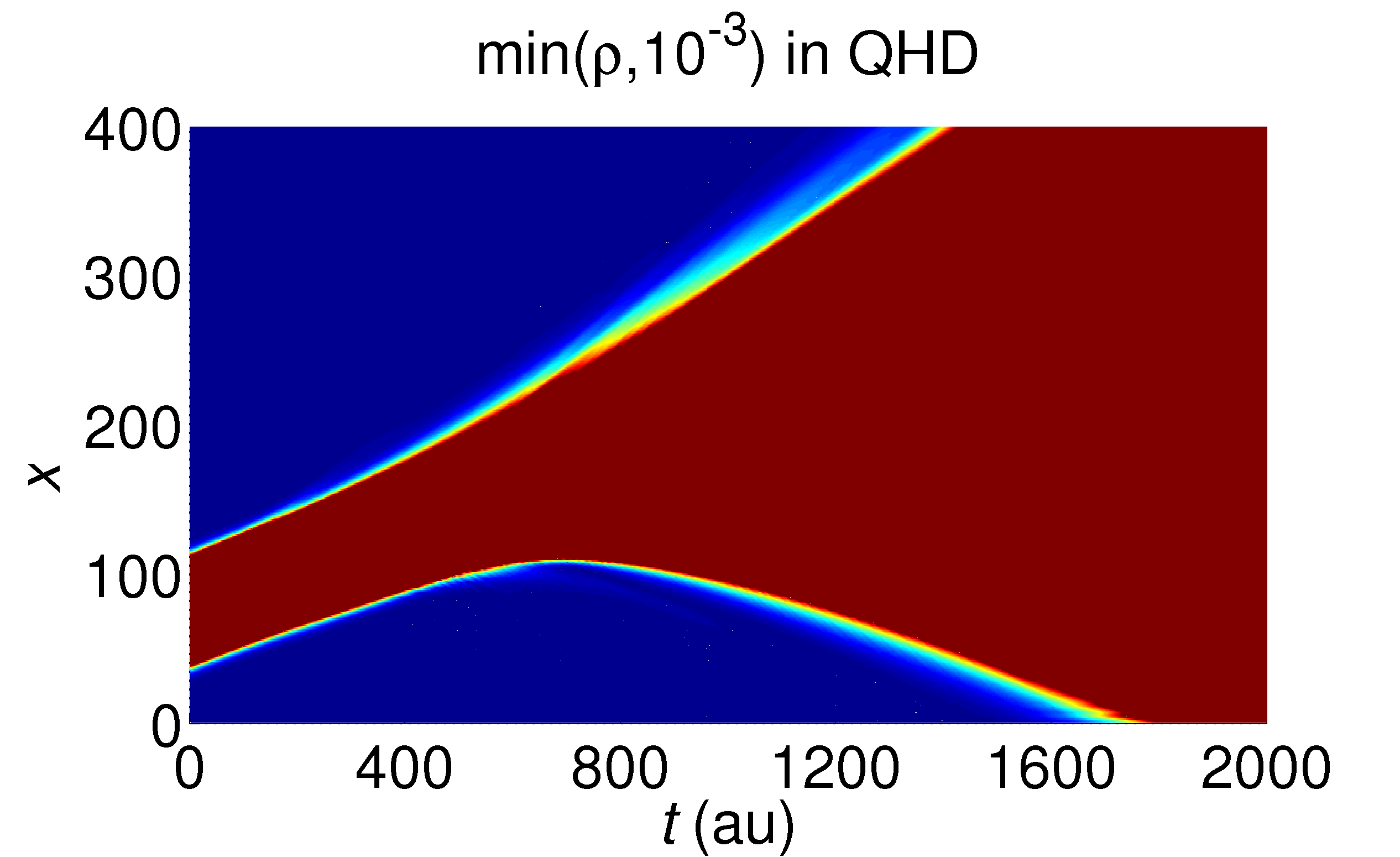}\includegraphics[width=6cm]{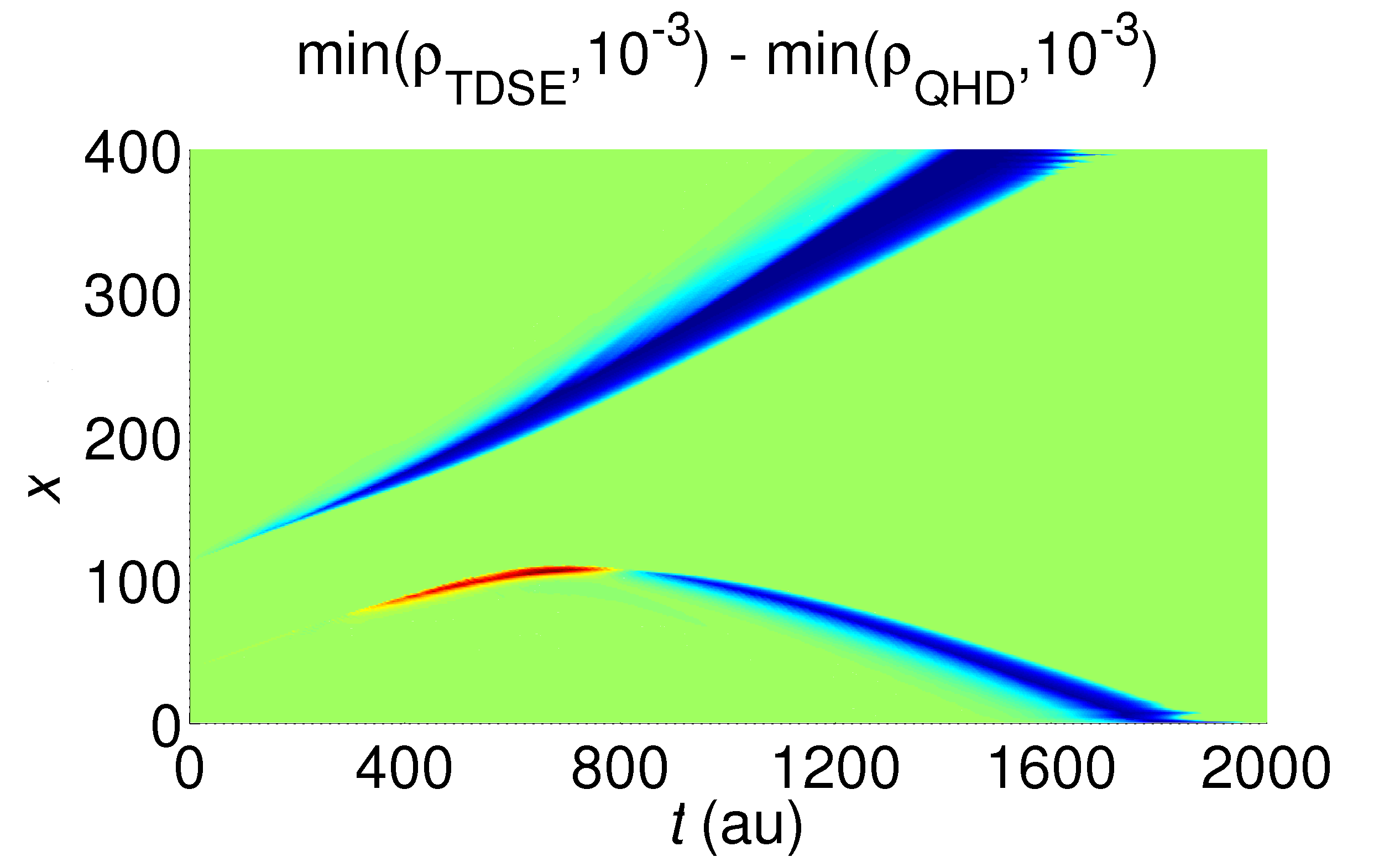}
\caption{We show the diffusive noise profile $\min(\rho_{\mathrm{QHD}},10^{-3})$ in the QHD solution, and the difference $\min(\rho_{\mathrm{TDSE}},10^{-3})-\min(\rho_{\mathrm{QHD}},10^{-3})$.  Here $x$ refers to the $x$-th meshpoint.}
\label{fig:noise}  
\end{figure}

The background ambient value $\rho_{A}$ is required in order to satisfied the mathematical and numerical well-posedness of the system such that the behavior of the system is not perturbed away from its proper character by compounding residual behavior, as shown in \textsection{3}.  Furthermore, from a phenomenological point of view, this value is nonrestrictive and physically easily justified -- for example, for a chemical reaction occurring in a solvent bath, or, similarly, any process occurring away from vacuum.

The discretization proceeds as in section \textsection{2} and \textsection{4}, where we adopt the local Lax-Friedrich's inviscid flux with van Leer's MUSCL slope limiting scheme.  Next we implement a standard explicit Runge-Kutta time discretization (see \cite{Cockburn1,SO} and \cite{MSEV}, or \cite{Michoski} for explicit details). 

Now we solve the resultant system using for our initial data (\ref{init1})-(\ref{init2}) explicitly that $\mu=0.16,\alpha=2,x_{0}=3$ and $x_{1}=6$, such that,  \[\rho_{0}= 10^{-10}+\left(\frac{1}{\sqrt{2\pi\mu}}\right)e^{\frac{-(x-3)^{2}}{0.32}}\quad\mathrm{and}\quad u_{0} = \left(2 V_{0}\right)^{1/2},\] with potential: \[V(x)=V_{0} \ \sech^{2}\left(\frac{1}{2}(x-6)\right).\]   It is worth noting that we have thus chosen a kinetic energy which is in the context of a mixed classical-quantum regime; which is just to say that some classical trajectories trasmit over the barrier, in addition to those that tunnel quantum mechanically.  For boundary data we use the approximate well-posed Dirichlet conditions discussed in \textsection{3}: \[\rho_{b} = \rho_{A}=10^{-10} \quad\mathrm{and}\quad u_{b} = 0.\]  We compare our solution to a finite difference scheme for the TDSE provided by Prof. Robert E. Wyatt \cite{Wyatt} in order to test the accuracy of our formulation.  The TDSE has equivalent initial settings, while the boundary conditions are given naturally via $\psi_{b}=\psi_{i,b}$ as discussed in \textsection{3}.  
\begin{figure}[!t]
\centering
\includegraphics[width=10cm]{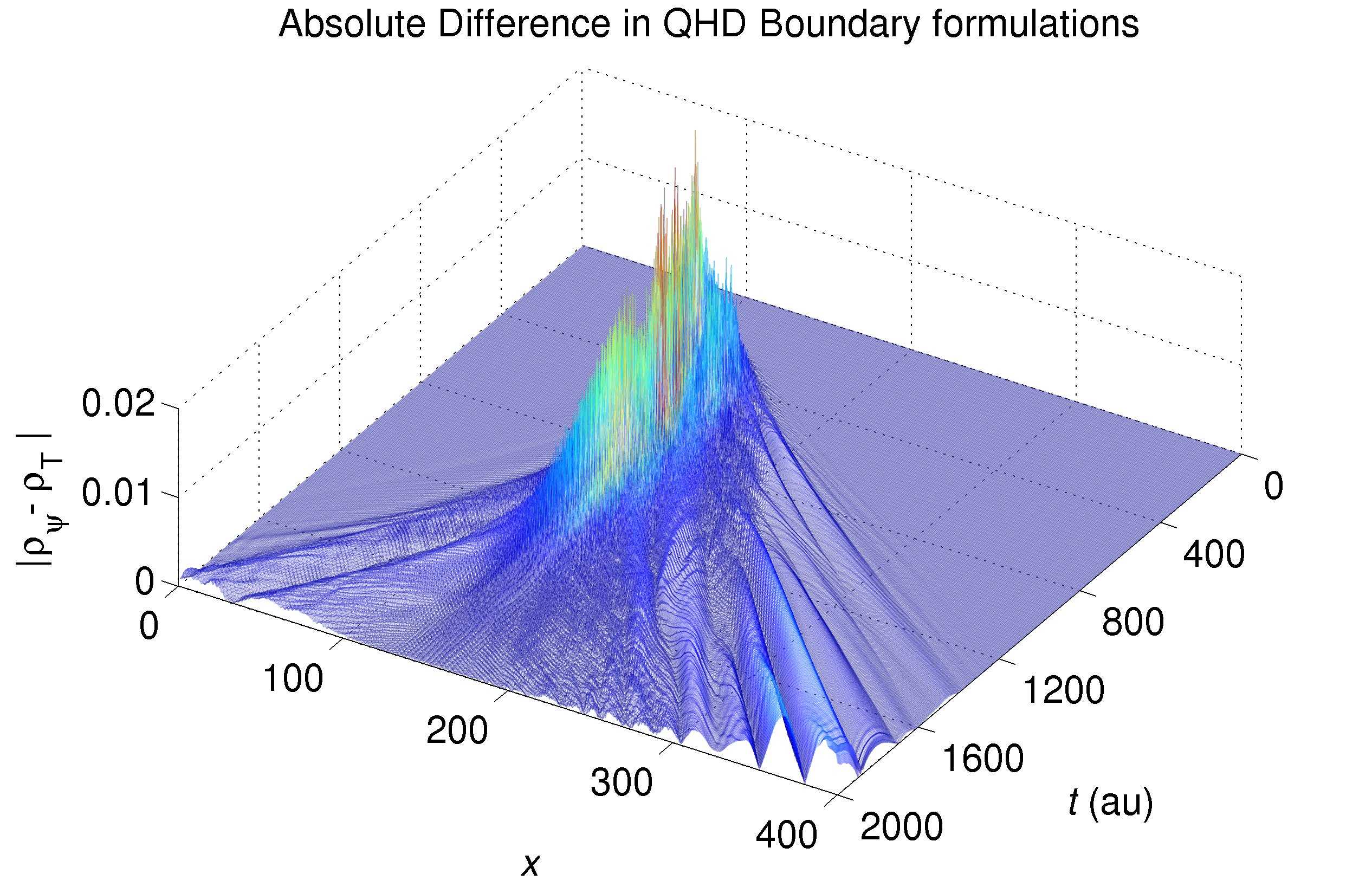}
\caption{We show the absolute difference between the QHD solution using the approximate boundary data from Figure \ref{fig:class} denoted $\rho_{\psi}$ with the transmissive boundary formulation from (\ref{transmissive}) denoted $\rho_{T}$.  Here $x$ refers to the $x$-th meshpoint.}
\label{fig:absdiff}  
\end{figure}

In Figure \ref{fig:class} these two solutions are compared.  Is is clear that the two solutions have the same qualitative behavior.  However they do show fundamentally different quantitative behaviors.  Analysis has shown that the two most prevalent sources of error between these two solutions are diffusion and boundary oscillations.  The boundary oscillations clearly occur due to the approximations discussed in \textsection{3}.  The diffusion, on the other hand, is a signature of the slope limiter in the QHD formulation and is shown in greater detail in Figure \ref{fig:noise}.  Here we confirm that the MUSCL slope limiting scheme is adding a type of ``artificial diffusion'' to the QHD solutions.  We have found that choosing a less restrictive slope limiter, such as the flux limiter of Osher presented in \cite{Osher}, does stably reduce the diffusion in our solutions.
 
We may now recover trajectories, or characteristics, of the solution by using the fact that (\ref{mass}) is satisfied at every time step (note that we show the alternative method of integrating velocity ``pathlines'' in \textsection{6}).  We may think of this equation as a kind of ``conservation of density'' here, and thus we simply employ Reynold's transport theorem (RTT): \begin{equation}\label{rtt}\frac{\partial}{\partial t} \int_{\tilde{\Omega}(t)} \rho dx + \int_{\tilde{\Gamma}(t)} \rho u_{rel} \cdot n dx = 0,\end{equation} where $u_{rel}$ is the relative velocity of the fluid with respect to the moving boundary $\tilde{\Gamma}(t)$.  First consider the case when $u(a) \approx 0$ such that we may choose $\tilde{\Omega}(t) = (a, y(t))$ where $y(t)$ is the moving boundary treated as an unknown.  By assumption and construction, $u_{rel}(a) = 0$, whereas for a trajectory we require $u_{rel}(y) = 0$.  Then integrating (\ref{rtt}) in $t$ we find \begin{equation}\label{trajectories}\int_a^{y(t)} \rho dx = \int_a^{y(0)} \rho dx.\end{equation} Let us define for each trajectory $y(t)$ with $y(0)=y_{0}$ the ``locally accumulated mass'' $M$ by: \[M(y_{0},t) = \int_{a}^{y(t)}\rho(x,t)dx.\]  Approximating each trajectory then directly follows from the equation $M(y_{0},t)=M(y_{0},0)$. 

\begin{figure}[!t]
\centering
\includegraphics[width=6cm]{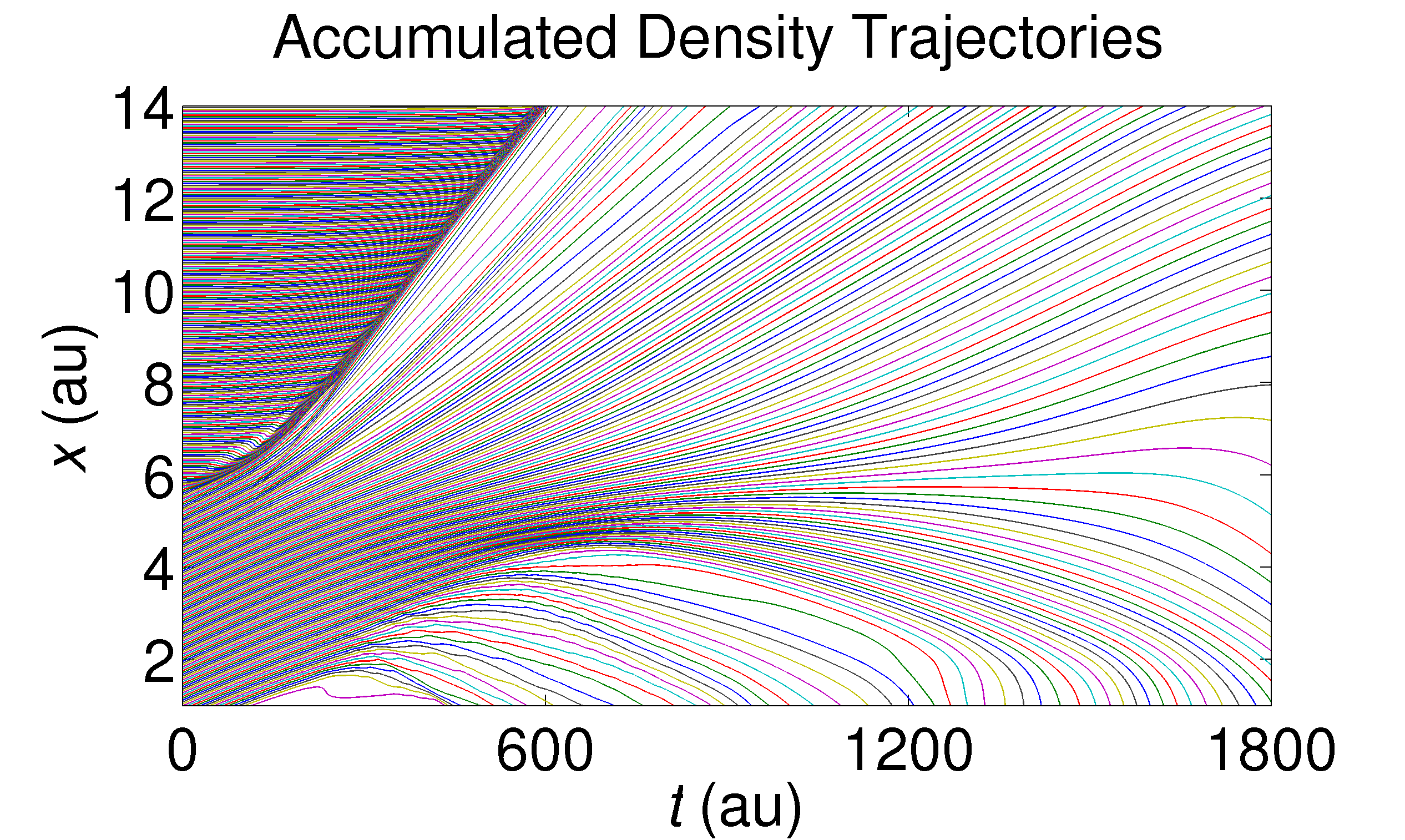}\includegraphics[width=6cm]{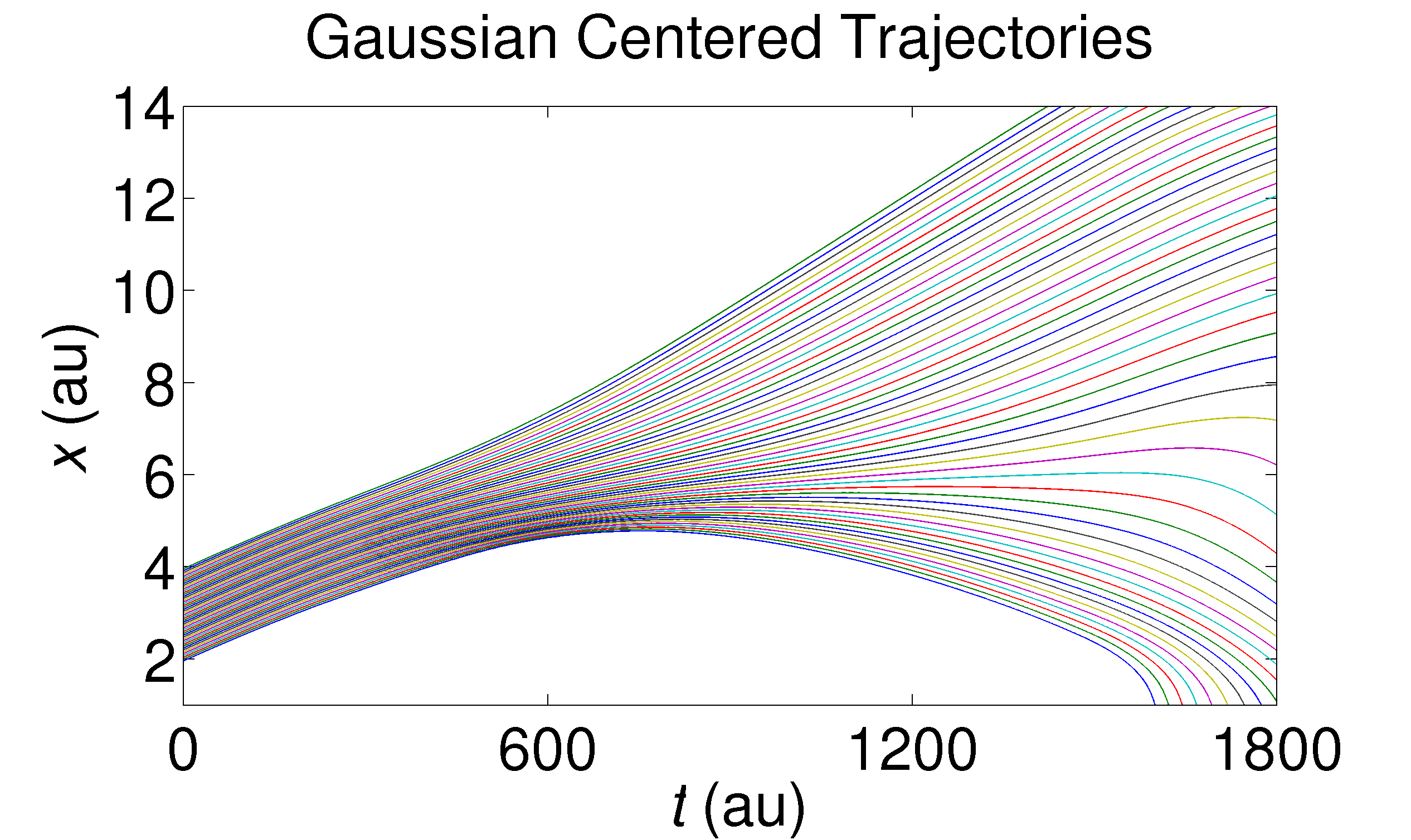}
\caption{We solve the accumulated density trajectories from (\ref{trajectories}) using the transmissive solutions from $\rho_{T}$ in Figure \ref{fig:absdiff}.}
\label{fig:AccTraj}  
\end{figure}
 
To continue let us denote $M_{i}(t)=M(x_{i},t)$, where $x_{i}$ is the $i$-th meshpoint. To compute $y(t)$, we compare $M(y_{0},t)$ to the increasing sequence $\lbrace M_i(t) \rbrace_{i=0 \ldots N}$ and find $j$ such that $M_{j-1}(t) \le M(y_{0},t) < M_{j}(t)$, which gives us that $y(t) \in [x_{j-1},x_j)$. Then to find $y(t)$ recall that we have from (\ref{shapefunctions}) an expansion
\[
\rho_h(t,x) = \sum_{l=0}^d c_l(t)N_l^j(x), \quad\mathrm{for}\quad x \in (x_{j-1},x_j)
\]
where the $c_l=c_l(t)$ are constants for every fixed $t$ and the shape functions $N_l^j(x)$ in our implementation are translated versions of polynomials $\{P_l\}_{l=0}^{d}$ on $[-1,1]$. That is using $f_{j}\colon[x_{j-1},x_{j}]\mapsto[-1,1]$ where \[f_j(x)=2\left(\frac{x-x_{j-1}}{x_j-x_{j-1}}\right)-1,\] we find, $N_l^j(x) = P_l( f_j(x) )$.  Then solving for $y(t)$, formulated via
\[
M(y_{0},t) = M_{j-1}(t) +\int_{x_{j-1}}^{y(t)} \rho_h(x,t) \mathrm{d}x = M_{j-1}(t)+ \int_{x_{j-1}}^{y(t)} \sum_{l=0}^d c_lP_l( f_j(x) )  \mathrm{d}x,
\]
can be recast by a change of variables, as solving for $X$ in 
\begin{equation}\label{root}
M(y_{0},t) = M_{j-1}(t) + \left(\frac{2}{x_j-x_{j-1}}\right) \int_{-1}^X \sum_{l=0}^d c_lP_l(z)dz,
\end{equation}
after substituting $z=f_j(x)$.  But that just corresponds by the change of variables, to \[X = 2\left(\frac{y(t)-x_{j-1}}{x_j-x_{j-1}}\right)-1.\]  Then a solution to $X$ exists by the intermediate value theorem, and since the integrand is positive it is uniquely determined as the only solution on $[-1,1]$ to the polynomial equation of degree $d+1$ arising from (\ref{root}). We my then, for example, in the piecewise linear case (i.e. $d=1$) use the quadratic formula to recover $X$ and hence the position of $y(t)$ within $\mathcal{G}_{j}$.

\begin{figure}[!t]
\centering
\includegraphics[width=10cm]{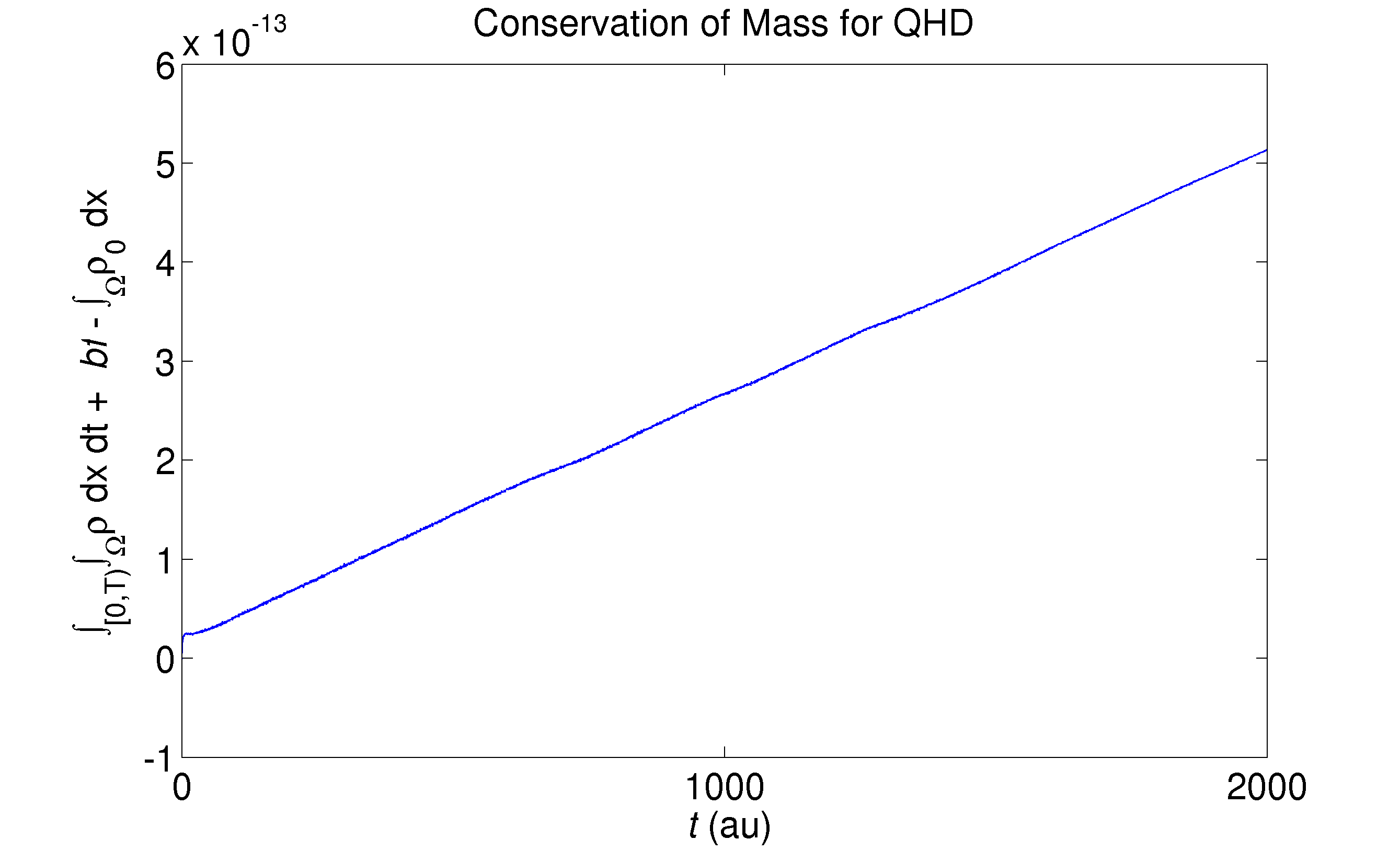}
\caption{Here we show mass conservation in the QHD regime given transmissive boundaries, where $bf=\int_{[0,T)}\int_{\partial\Omega}\rho dxdt$ is the boundary flux.}
\label{fig:CM}  
\end{figure}

Similarly, we also work in the other direction, with the balance of the mass in $[y(t),x_{j}]$ such that the analogous integral equation becomes:
\[
\int_{Y}^{1} \sum_{l=0}^d c_l P_l(z)dz,
\]
which provides for a consistency check on the accumulated density in either direction.  Consequently we have that the sequence $\lbrace y(t) \rbrace_{t=1,\ldots,T}$ provides a numerical approximation to the position of a particle initially at $y_{0}$ when $t=0$ at our given set of later times.  

This formulation holds as long as our hypothesis, $u(a)\approx 0$, is satisfied.  However, we can immediately extend this result to include the case $u(a)\neq 0$.  That is, after integrating in $t$ we note that (\ref{rtt}) becomes:
\[\int_a^{y(t)} \rho dx = \int_a^{y(0)} \rho dx + \int^t_0 \rho u(a) dt.\]
This gives us an alternative equation to find $y(t)$, where we must only add $\int^t_0 \rho u(a) dt$ to the accumulated density $M$ at every $t$.  We further note that this basic framework may also be adapted to higher dimensions (see \cite{Price}).

Now, we again solve our system with (\ref{trajectories}) using for (\ref{init1})-(\ref{init2}) that $\mu=0.16,\alpha=2,x_{0}=3,x_{1}=6$, however now we introduce the transmissive boundary condition: \begin{equation}\label{transmissive}\boldsymbol{U}_{h}^{n}|_{\mathcal{K}_{ji}}=\boldsymbol{U}_{h}^{n}|_{\mathcal{K}_{ij}},\end{equation} as discussed in \textsection{3}.  In Figure \ref{fig:absdiff} it is clear that the behavior between the solutions with transmissive and approximate solutions is quite distinct, and that boundaries are, so to speak, felt in the interior solution even before significant density has reached $\partial\Omega$.  We use the transmissive boundary conditions to construct the accumulated mass trajectories derived above, as they seem to represent more physically cogent boundaries.  The results are shown in Figure \ref{fig:AccTraj}, where the the ``Gaussian centered trajectories'' are simply the trajectories containing the majority of the initial density; that is, those trajectories whose initial positions are $\pm 1$ au from the center of $\rho_{0}$.

\begin{figure}[!t]
\centering
\includegraphics[width=6cm]{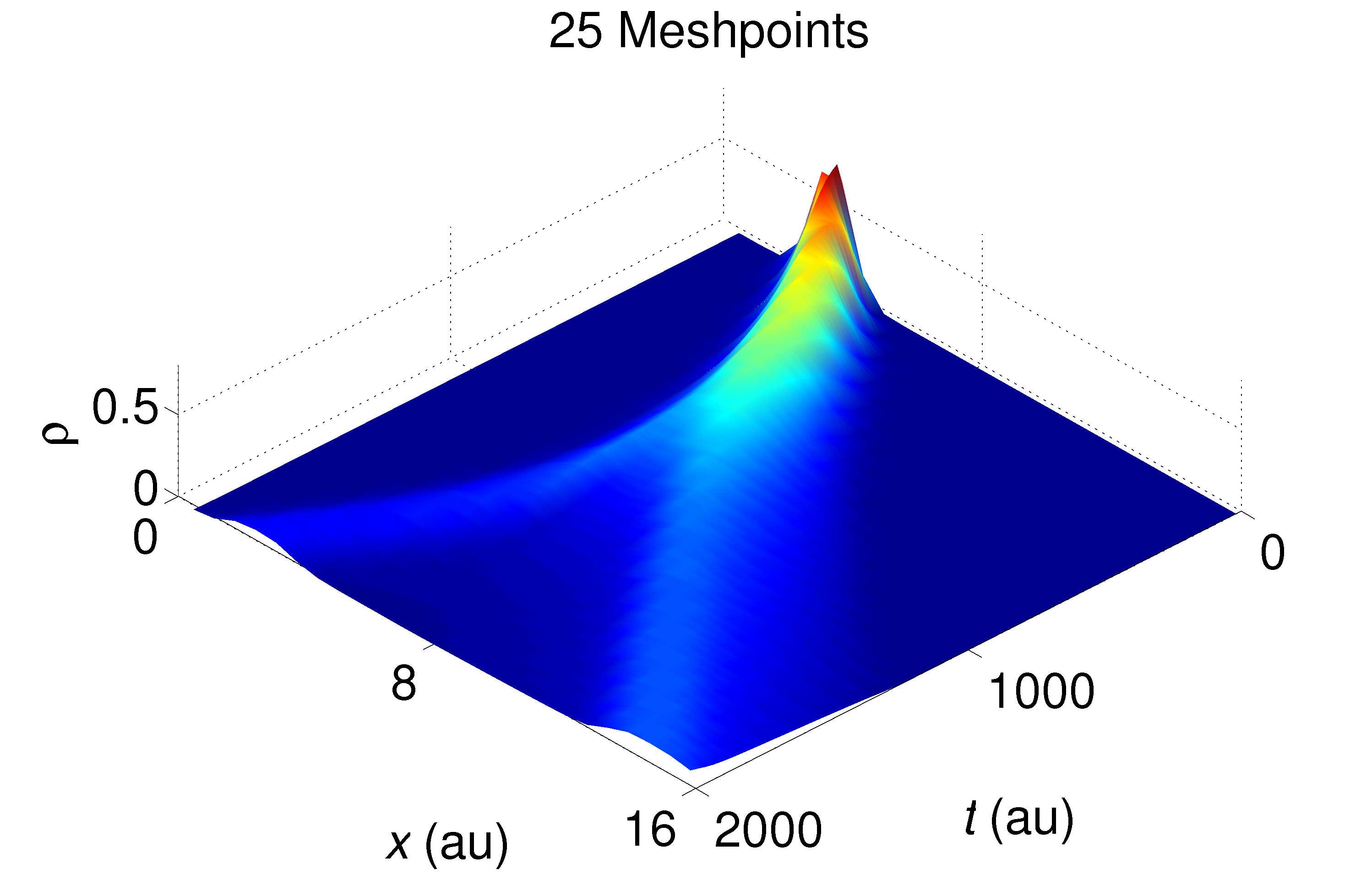}\includegraphics[width=6cm]{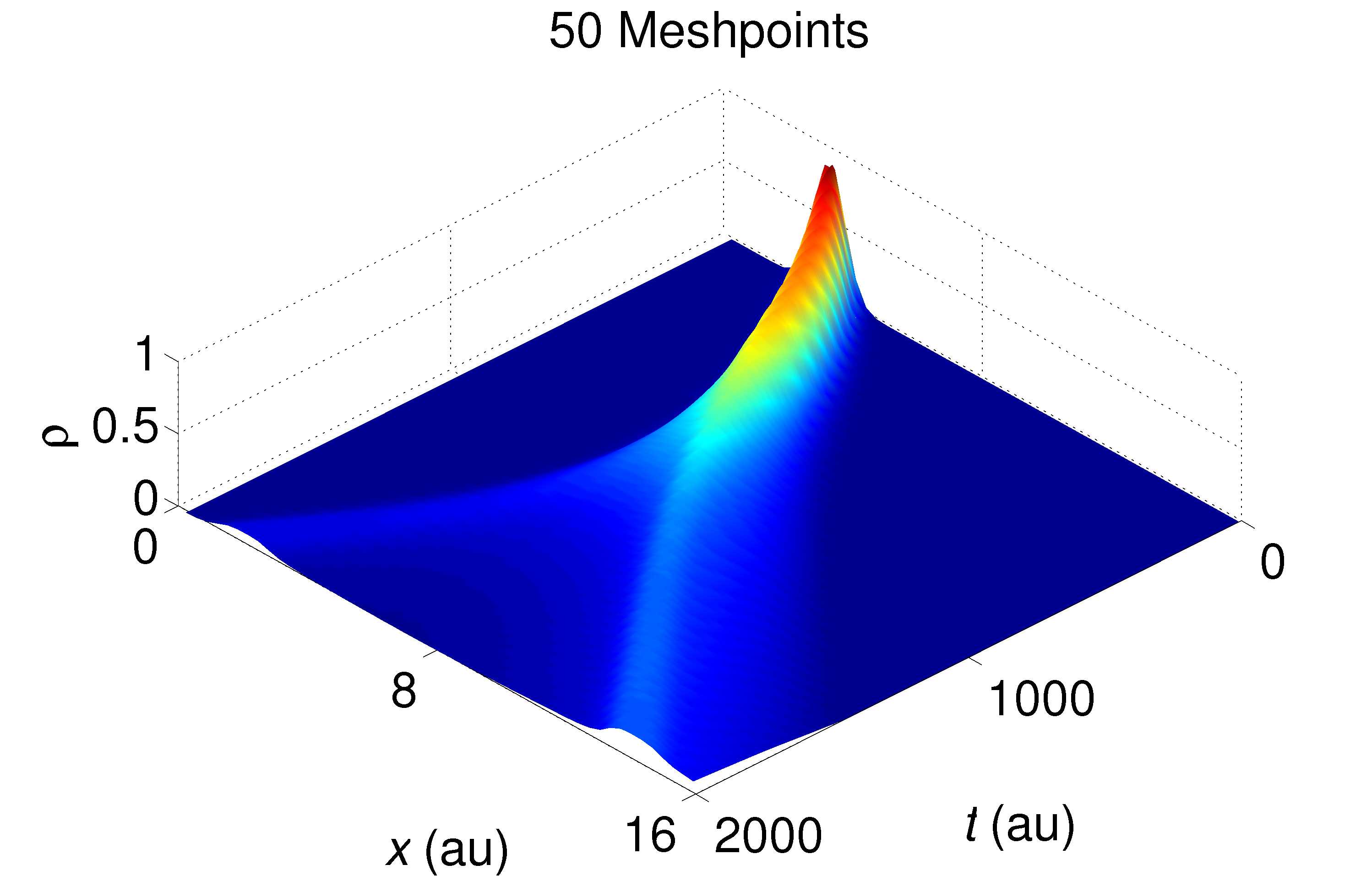}
\caption{Here we show the remarkable spatial invariance of the solution.  These represent the same solution as that given in Figure \ref{fig:class}, except the left graph is with 25 meshpoints and 100 timesteps, and the right at 50 meshpoints and 200 timesteps.}
\label{fig:invar}  
\end{figure}

We further show that the MDG method is a conservative scheme.  That is, in Figure \ref{fig:CM} we show that the density is effectively (numerically) normalized to one on the domain, when taking into account the boundary flux.  That is, we see linear error growth at machine precision over 10,000 timesteps.  Another feature of the solution which is attractive in the sense of practical applications, is that the spatial invariance demonstrated by the solutions.  In Figure \ref{fig:invar} we show the this feature, where the same calculation from Figure \ref{fig:class} is graphed, where there 400 meshpoints and 10,000 timesteps were used in order to compare with the TDSE.  However, as is clear from Figure \ref{fig:invar}, with only 25 meshpoints the solution provides the same qualitative answer.  This is an important feature in chemical applications where computations must scale in $3N$ dimensions, for $N$ the number of atoms in the molecular system of interest (see for example \cite{MCFHV}).

\section{\texorpdfstring{\protect\centering $\S 6$ Recovering $\psi$ and $S$ in both frames}{\S 6  Recovering psi and $S$ in both frames}}

\begin{figure}[h]
\centering
\includegraphics[width=6cm]{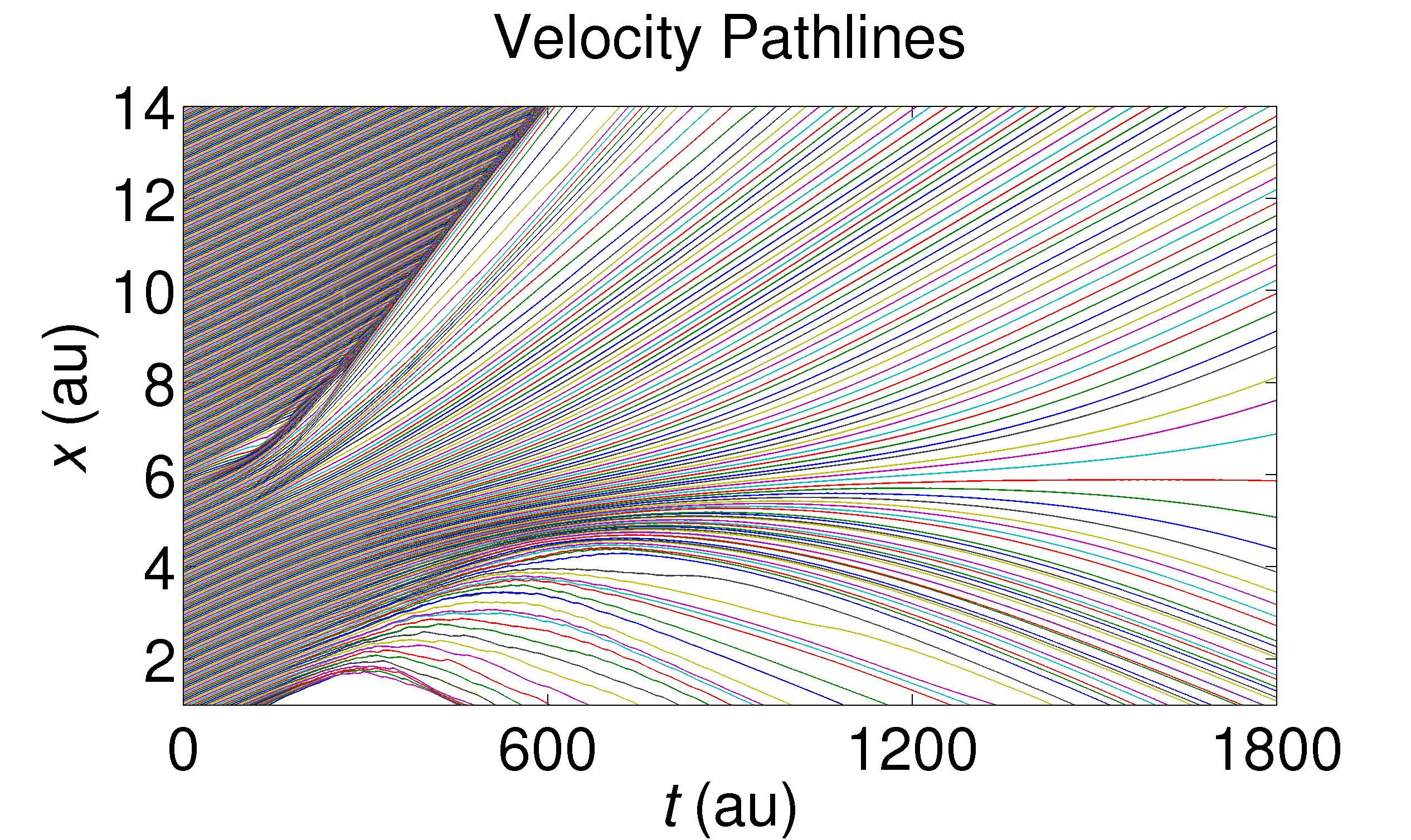}\includegraphics[width=6cm]{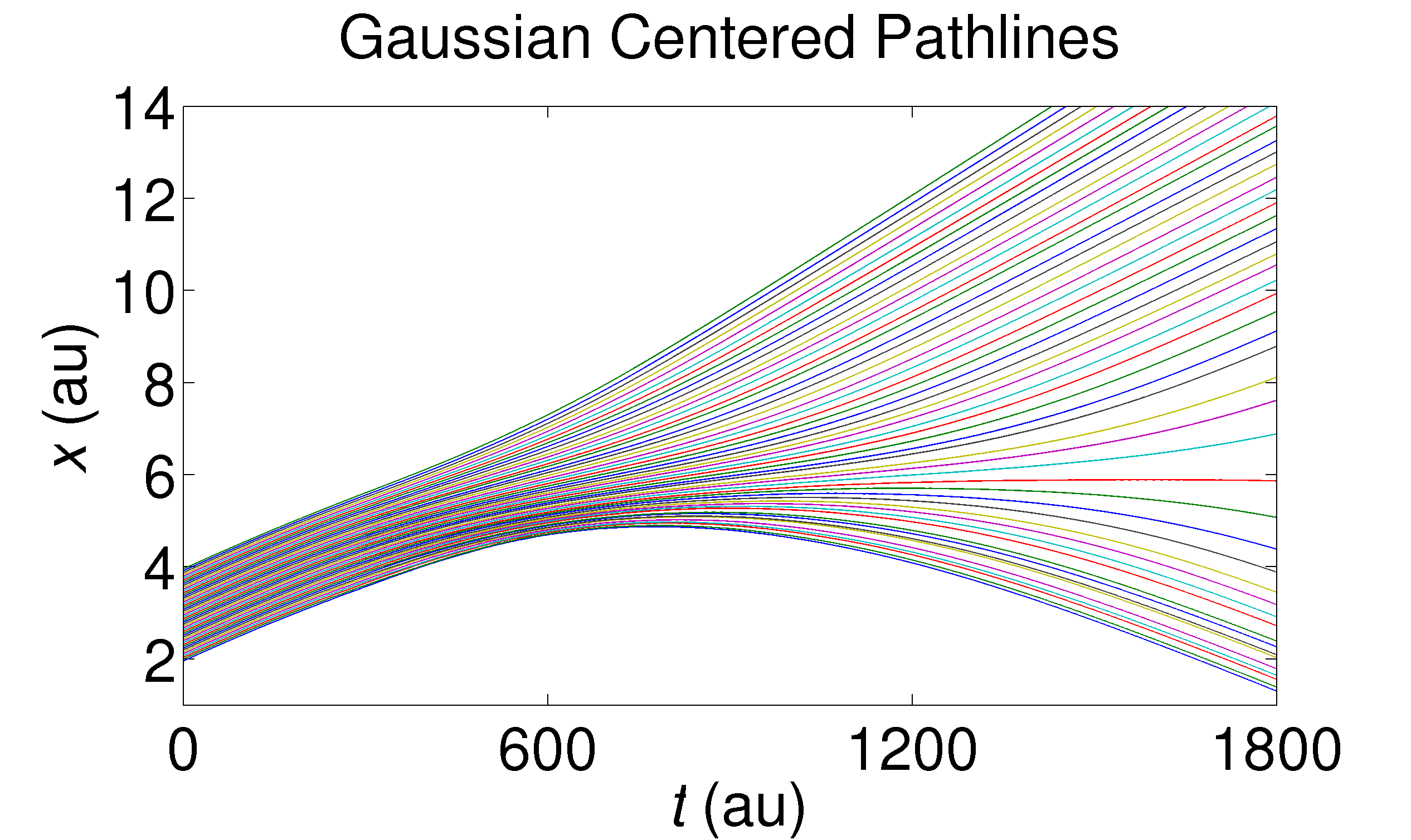}
\caption{We graph the quantum trajectories using (\ref{veltraj}) to solve for $\vec{r}$, which can be compared with the accumulated mass trajectories shown in Figure \ref{fig:AccTraj}.}
\label{fig:veltraj}  
\end{figure}

Now that we have solutions in the Eulerian and Lagrangian coordinate frames as given in \textsection{4} we may recover the important variables $\psi$ and $S$ in either frame.  First we note that we may alternatively recover the trajectories using the solution $\boldsymbol{U}$ from \textsection{4} to solve the initial value problem: \begin{equation}\label{veltraj}\frac{d\vec{r}}{dt}=\boldsymbol{u}(t,\vec{r}) \quad \mathrm{with}\quad \vec{r}_{|t=0}=\vec{r}_{0}.\end{equation}  We recover these $\vec{r}$ by direct integration, and compare them to those computed via (\ref{trajectories}) (see figure \ref{fig:veltraj}),  where we refer to the  $\vec{r}$ trajectories computed in (\ref{veltraj}) as the ``velocity pathlines.''

\begin{figure}[!b]
\centering
\includegraphics[width=6cm]{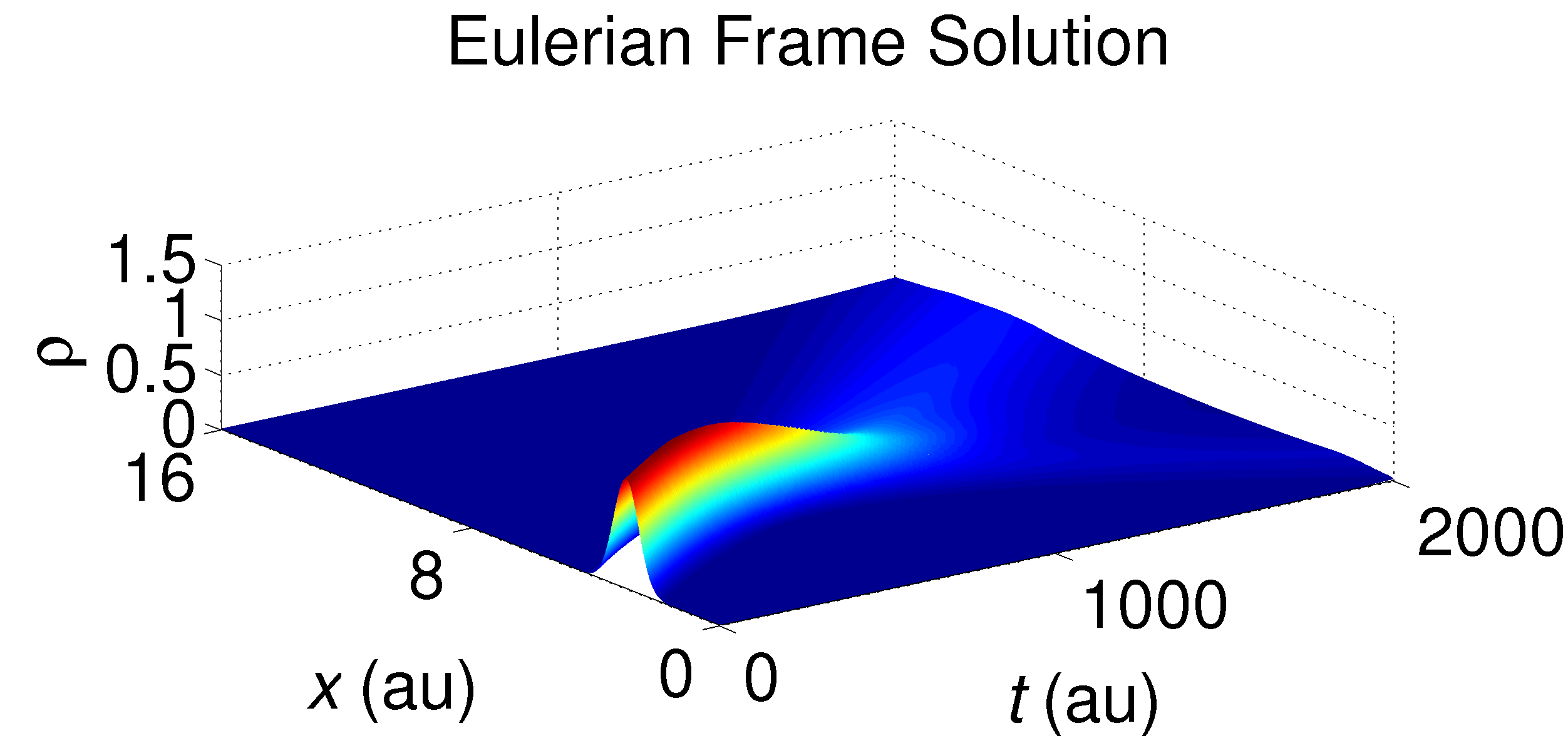}\includegraphics[width=6cm]{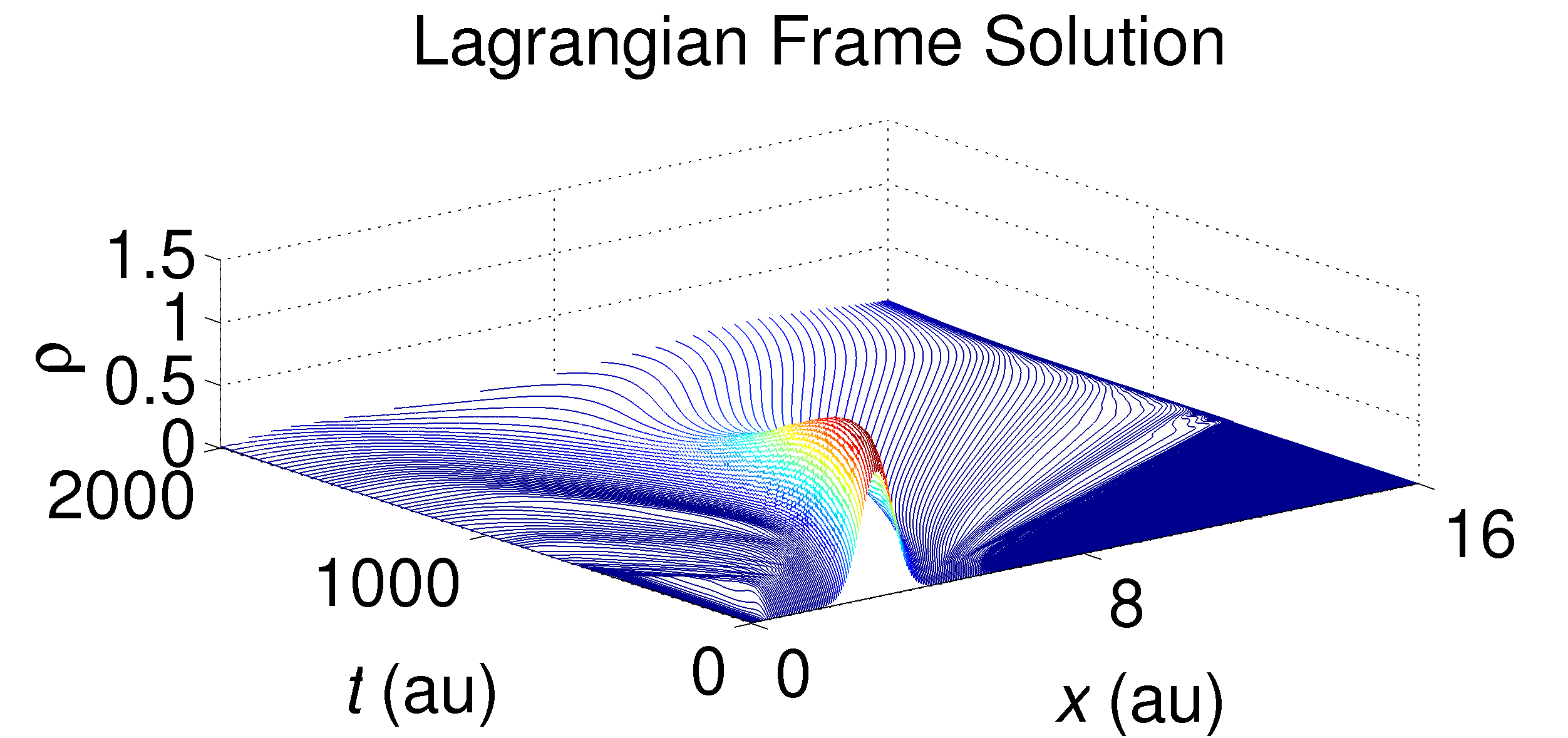}
\caption{The Eulerian solution $\rho(t,x)$ and the corresponding Lagrangian solution $\rho(t,\vec{r})$ for the same initial condition settings as in figure \ref{fig:AccTraj} using the conservation form of the trajectories (\ref{trajectories})..}
\label{fig:framerho}  
\end{figure}

The trajectories computed using the velocity field (\ref{veltraj}) are shown in Figure \ref{fig:veltraj} and show qualitatively similar behavior to the trajectories computed using the accumulated mass formulation in (\ref{trajectories}). There is no necessarily unique way of arriving at the trajectories one chooses to represent the solution in the Lagrangian frame.   For example, one may utilize a method which weights the solutions between (\ref{trajectories}) and (\ref{veltraj}).  That is, we may compute the trajectory positions via (\ref{veltraj}) and then offset these by a weighted average of the density conservation in (\ref{trajectories}).  We provide details on particular alternative in appendix A and show an example case.

It is now possible to solve for a number of derived variables in either the Lagrangian or Eulerian frames in order to recover the phase information of the quantum wave-packet associated to each characteristic pathline.  First we recover the trajectory-wise solutions $\rho(s,\vec{r})$ and $\boldsymbol{u}(s,\vec{r})$, and then compute the variables: \begin{equation}\label{square} \nabla_{x}S = \sqrt{m} \boldsymbol{u} \quad \mathrm{and}\quad \psi = \sqrt{\rho} e^{i S / \hbar},\end{equation} where $S(s,\vec{r})$ is the \emph{quantum action} and $\psi(s,\vec{r})$ is the \emph{quantum wavefunction}.  Recall that $R(s,\vec{r})=\sqrt{\rho(s,\vec{r})}$ as shown in \textsection{3}, such that using $\boldsymbol{v}$ for the velocity from \textsection{2} we recover from (\ref{square}) the more familiar formulation: \begin{equation}\label{both}\nabla_{x}S = m \boldsymbol{v} \quad \mathrm{and}\quad \psi = R e^{i S / \hbar}.\end{equation} 

\begin{figure}[!t]
\centering
\includegraphics[width=6cm]{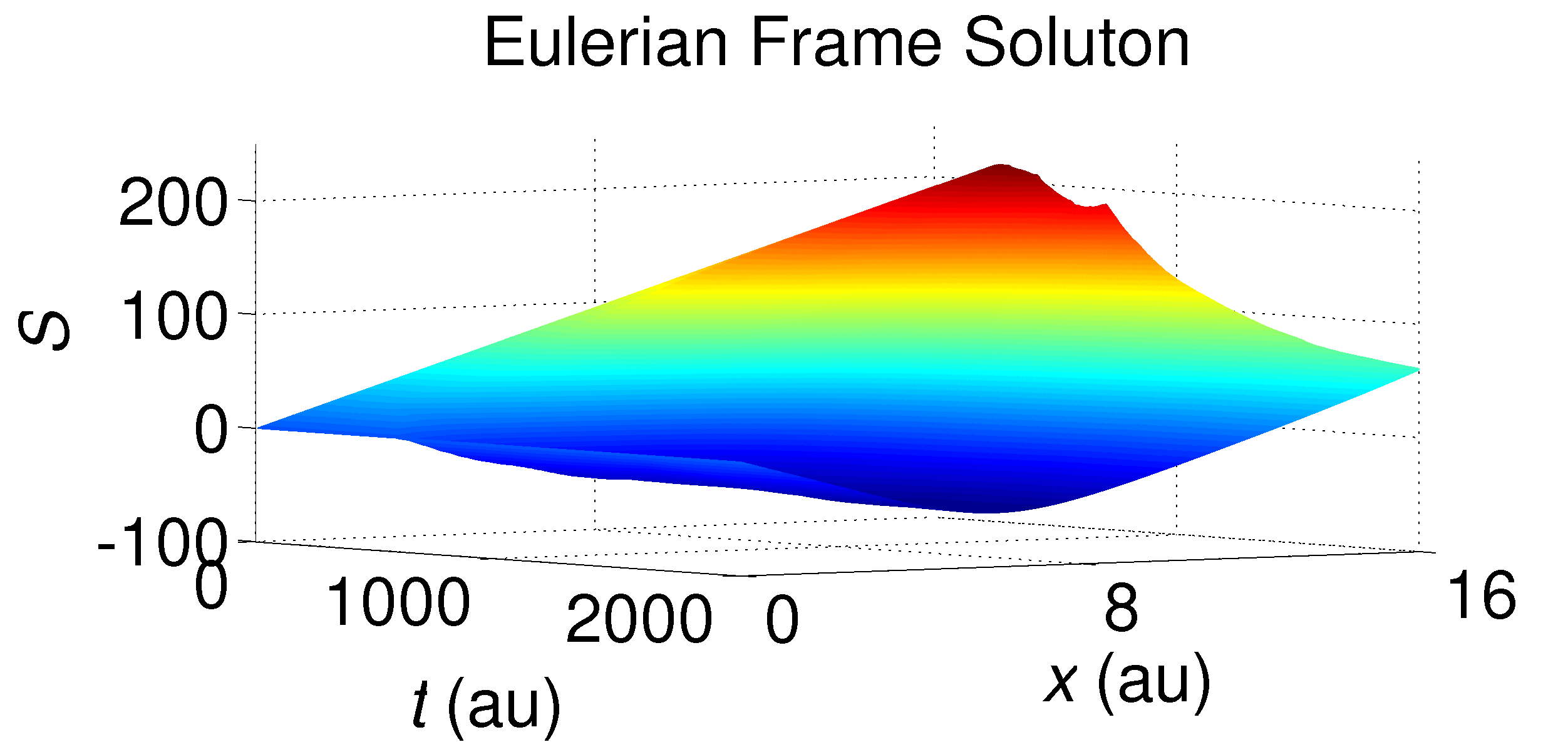}\includegraphics[width=6cm]{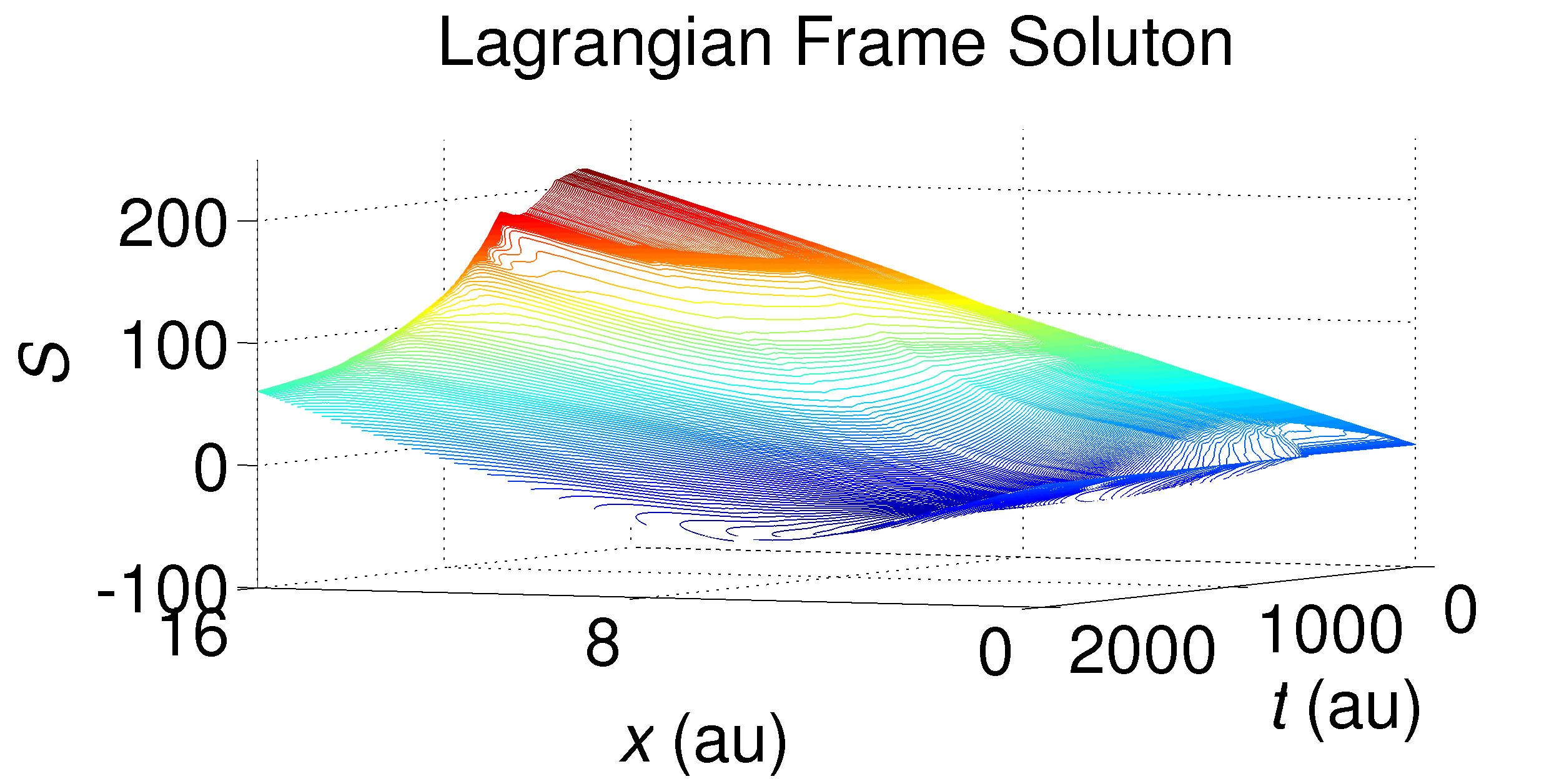}
\caption{A graph of the Eulerian solution $S(t,x)$ and the corresponding Lagrangian solution $S(t,\vec{r})$ for the same initial condition settings as in figure \ref{fig:framerho} using the conservation form of the trajectories (\ref{trajectories}).}
\label{fig:frameS}  
\end{figure}

It is important to note that up to a constant of integration, $S$ and $\psi$ are completely determined by the solution (\ref{aprox}).  Also, (\ref{both}) is satisfied in both reference frames, so we now have the following solutions: \begin{equation}\rho(s,x), \rho(s,\vec{r}), \boldsymbol{v}(s,x), \boldsymbol{v}(s,\vec{r}), \psi(s,x), \psi(s,\vec{r}), S(s,x), \ \mathrm{and} \ S(s,\vec{r}).\end{equation}  These solutions are graphed in figures \ref{fig:framerho}--\ref{fig:frameS}, where it is interesting to note that the two frames draw out different aspects of the solution. While the Lagrangian frame tracks individual ``particle'' trajectories across the function profiles, it misses some of the nuance in the continuous structure of the surface; which is naturally recovered by the Eulerian frame solution.  Furthermore, as lower resolution, we find that the conservation based trajectories from (\ref{trajectories}) are more well-behaved than the velocity based trajectories from (\ref{veltraj}).

\section{\texorpdfstring{\protect\centering $\S 7$ Conclusion}{\S 7 Conclusion}}

We have presented a numerical solution to the quantum hydrodynamic equations of motion as posited in the context of quantum hydrodynamics with chemical applications.  Our approximate solution is a rescaled (in time) version of the standard QHD equations and is the first model of its type presented in a mixed discontinous Galerkin framework in the context in which it arises in chemical applications. Our solution further shows good stability, up to a \emph{stiffness} of the system of equations which is a well-known feature of the QHD system of equations, and a scale invariance behavior which makes it very appealing for the so-called ``fast and dirty computations'' often needed in realistic chemistry applications.  Additionally we have shown in a rigorous and consistent way how to prescribe proper boundary data, which is often bypassed in the usual Lagrangian formulations of the system.  We have further demonstrated that in the conservation formulation of this system, the \emph{quantum wavefunction} $\psi$ and \emph{quantum action} $S$, which are used as motivation for the derivation of QHD systems to begin with (e.g. \cite{Madelung,Bohm2,Wyatt}), are in fact completely determined (up to a constant of integration) by the solutions $\varrho$ and $\boldsymbol{v}$.  

Finally, it is worth mentioning that these solutions are very closely related to quantum hydrodynamic solutions which have been extensively studied in other fields (see \cite{Gardner,Jungel,LM,CHM,JG}), but still maintain some important differences.  One of the most important and prohibitive aspects of the quantum chemical formulation of QHD, is that the potential surface $V$ arises from a multiple of $3$N degrees of freedom of each quantum subsystem, for N the number of atoms in each molecular subsystem (for example in an intramolecular rearrangement).  This arises from the interpretation of the wavefunction $\psi$ as being the foundational variable in the dynamics of the quantum subsystem in the chemical models.  Clearly, even for relatively small molecules, this immediately leads to extremely difficult numerical problems.  In this sense it is important to have a numerical scheme which is easily parallelizable, fast, robust and accurately reflects the mathematical character of the solution.  The MDG formulation presented herein is a numerical method that fulfills these requirements, and offers a viable solution to some of the many difficulties which arise in the complicated solution space of chemical quantum hydrodynamics.  The scale invariance of the solution makes it even an alternative approach to the Lagrangian formulation; up to the ``formal'' accuracy of solutions.  

\section{\texorpdfstring{\protect\centering $\S$ Acknowledgements}{\S  Acknowledgements}}

The authors would like to thank Bob Wyatt for his many illuminating discussions and insights into the world of chemical quantum dynamics, and for providing the TDSE code for comparative analysis.  The first author would further like to thank Bas Braams for his help understanding potential energy surfaces, Daniel Dix for his detailed explanation of Bohmian mechanics and the quantum Talbot effect, and to further express sincere gratitude to John Stanton for his support.  The second author was partially supported by the Department of Energy Computational Science Graduate Fellowship, provided under grant number DE-FG02-97ER25308.   The fourth author was partially supported by the NSF Grant DMS 0607953. 
  
\section{\texorpdfstring{\protect\centering $\S$ Appendix A}{\S  Appendix A}}

The conservation method of recovering trajectories in (\ref{trajectories}) and the velocity integration method of recovering trajectories in (\ref{veltraj}) in no way exhaust the number of ways of representing solutions in the Lagrangian frame.  In fact, there are an infinite number of ways of choosing trajectories.  We introduce a way of computing a subset of these, and refer to these as ``offset methods.''

\begin{figure}[ht]
\centering
\includegraphics[width=6cm]{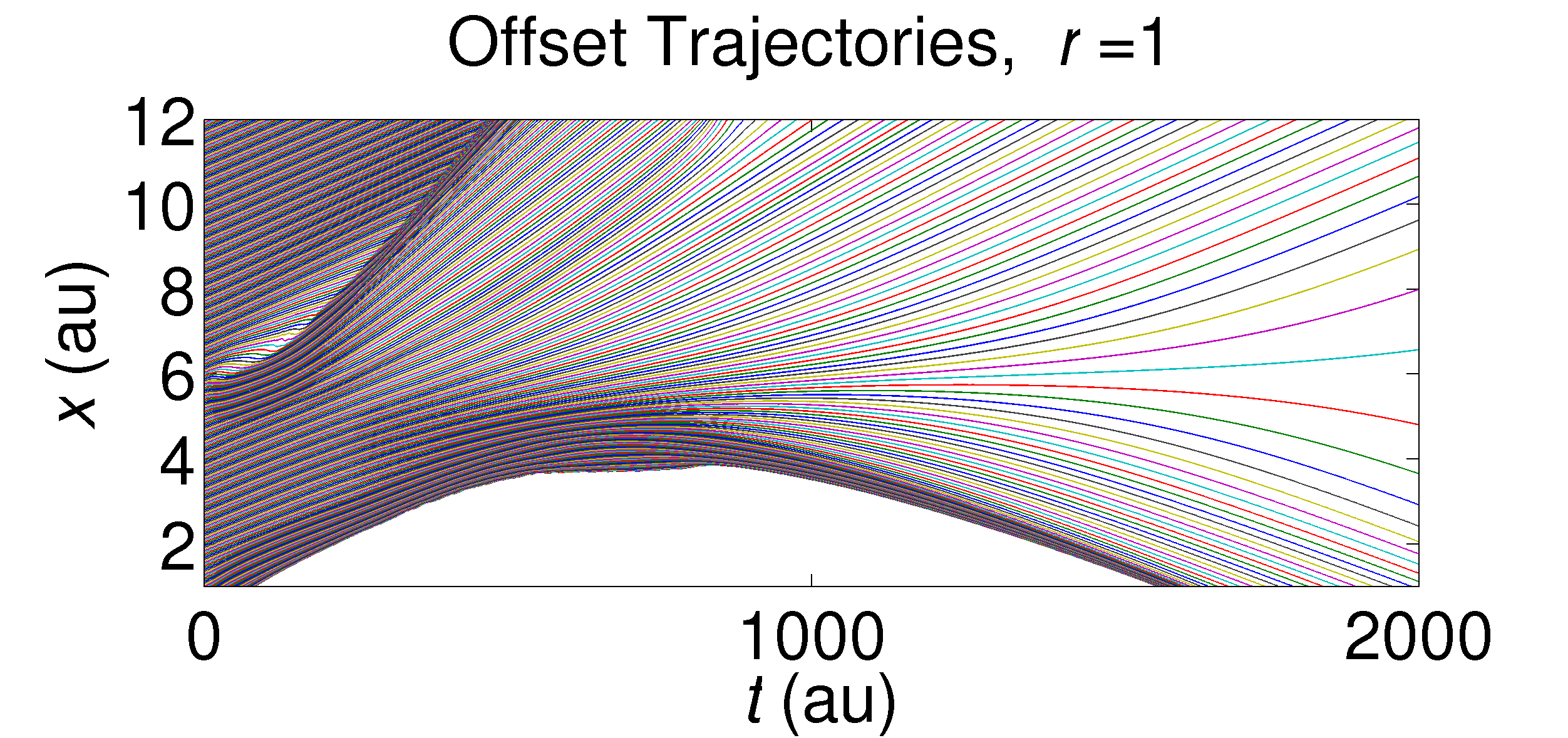}\includegraphics[width=6cm]{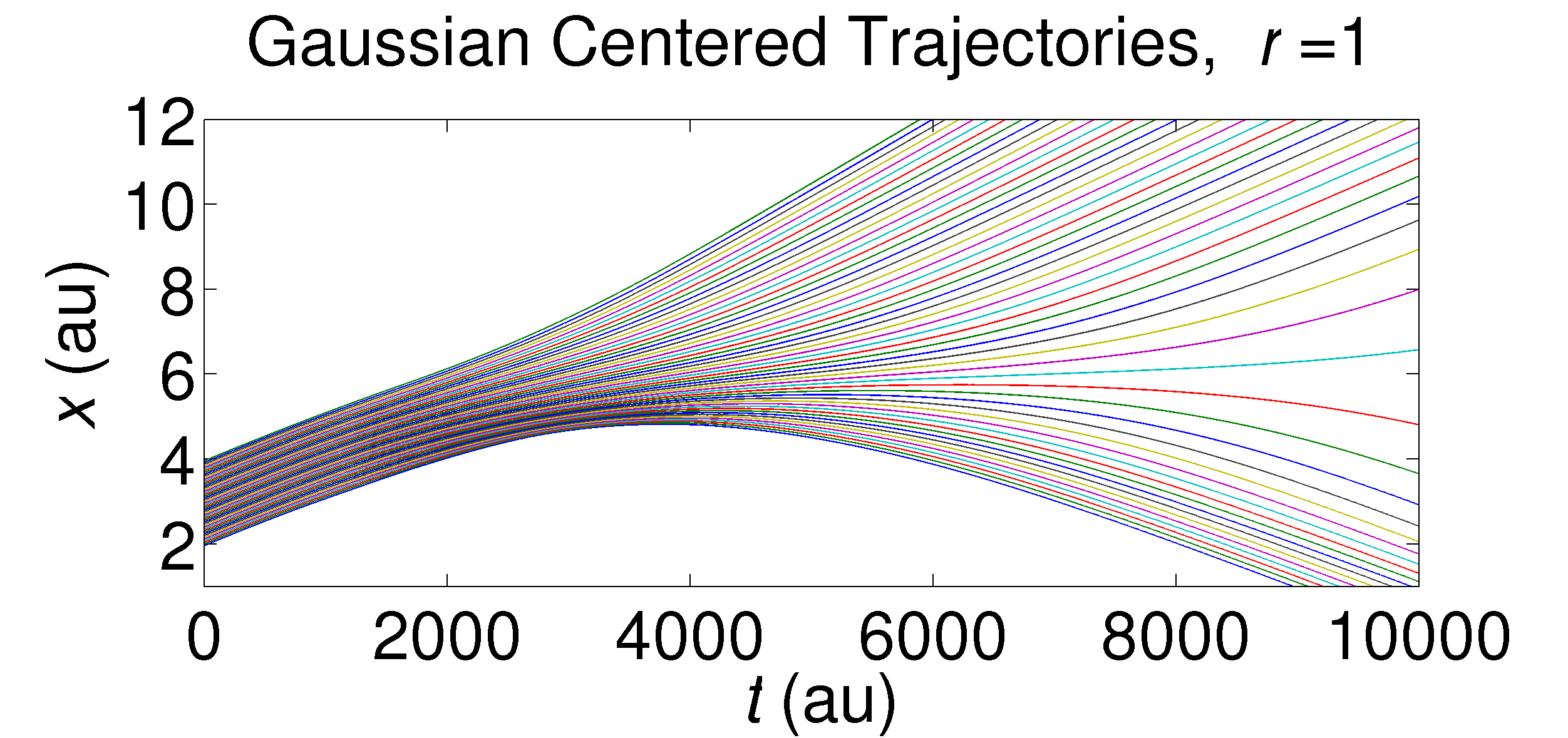}\\ \includegraphics[width=10cm]{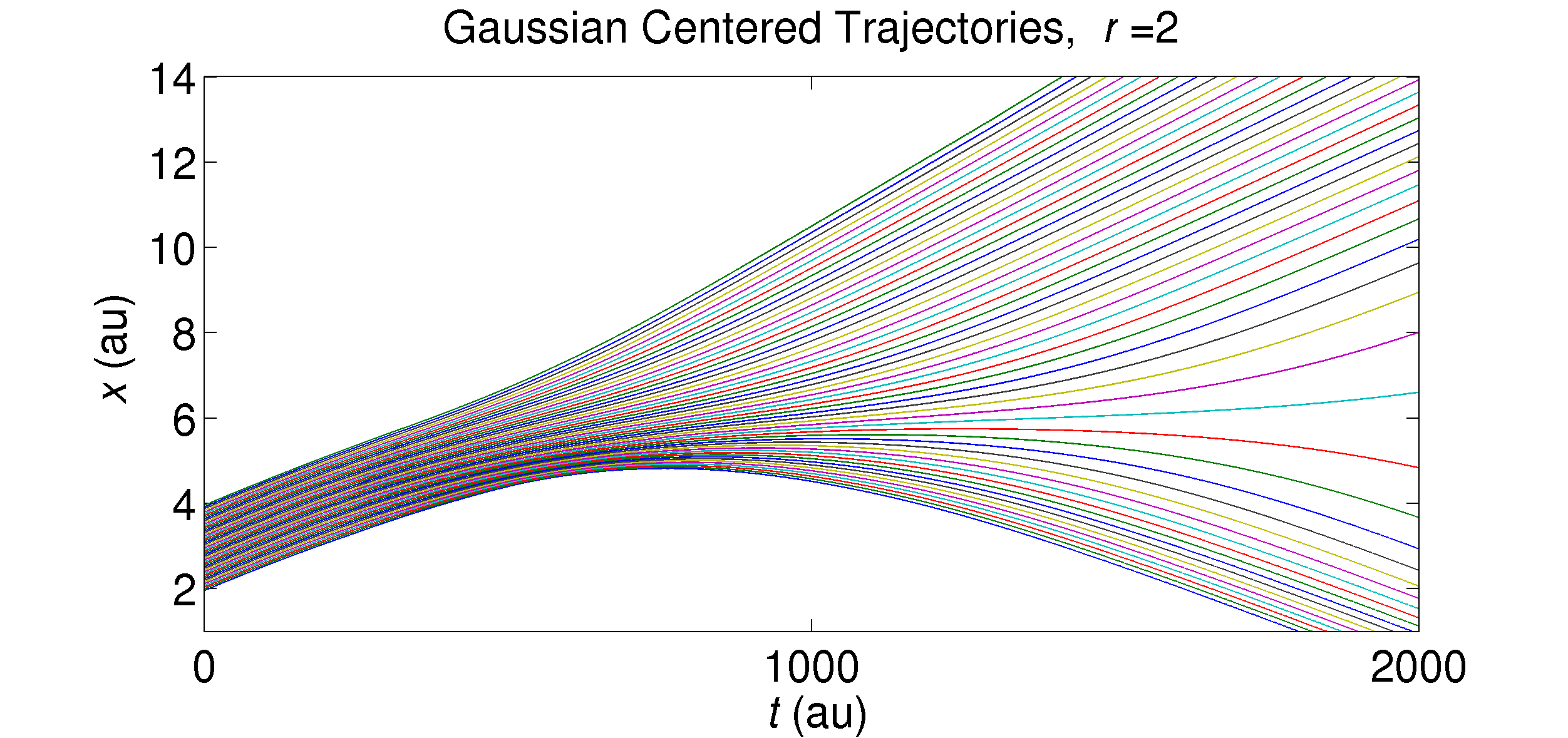}
\caption{On the top we show the quantum trajectories using the offset method solution of the same problem in Figure \ref{fig:veltraj} with $r=1$; and on the bottom we show the same trajectories using $r=2$.}
\label{fig:offsettraj}  
\end{figure}

The offset solution relies mainly on velocity integration but includes some information from mass conservation as follows: velocity integration provides an estimated position for each particle at the following time-step. Then one works through particle by particle, starting at the new estimated position and using mass conservation to estimate the new positions of its neighbors (a tunable number of consecutive elements on either side) offset from the velocity estimate of the `current' particle.  We set our tuning parameter to $r$ here on both sides, though there is no reason \emph{a priori} to choose a symmetric (with respect to either side) tuning parameter.  Generically this provides a set of estimates for the position of each particle: one directly from integration, and others via the relationship of that estimated position to the relative estimated position of its nearest neighbors. 

That is, if $P_m^m$ is the velocity estimated position, and $P_{m-r}^m$ and $P_{m+r}^m$ are the positions of the particles on either side that density conservation requires, and applying our symmetry constraint gives for the new position that: \[P_{\mathrm{new}} = w_{0}P_m^m+\sum_{i=1}^{r}\left(w_{i}P_m^{m-i}+w_{i}P_m^{m+i}\right),\] where the $w_{i}$'s are the weights for each component, in our examples computed with a Gaussian weighting functions $\omega_{i} = e^{-(\ln(2)/r^2)i^2}$ such that: \[w_{i} = \omega_{i}\Big/\sum_{i=0}^{r}\omega_{i}\quad\mathrm{for} \ i = 0,\ldots,r.\]  Then for $r=1$ we have $w_{0} = 1/2$ and $w_{1} =1/4$.  We show two examples of obtained offset trajectories in Figure \ref{fig:offsettraj}, which are located at distinct locations in the solution space.  Also note that these trajectories behave substantially different than those in Figures \ref{fig:AccTraj} and \ref{fig:veltraj}.

\end{document}